\documentclass[11pt,drafunctiontcls,onecolumn]{IEEEtran}

\usepackage{%
	amsmath,%
	accents,%
	amssymb,%
	amsthm,%
	graphicx,%
	array,%
	bchart,%
	caption,%
	cite,%
	diagbox,%
	epsfig,%
	latexsym,%
	multirow,%
	pgfplots,%
	sidecap,%
	subcaption,
	tikz,%
	varwidth,%
	wrapfig,%
	xcolor,%
	arydshln,%
	slashbox%
}

\usepackage[inline]{enumitem}
\usepackage[hyphens]{url}
\usepackage[hidelinks]{hyperref}
\hypersetup{breaklinks=true}

\sidecaptionvpos{figure}{c}

\usetikzlibrary{%
	arrows,%
	automata,%
	calc,%
	patterns,%
	plotmarks,%
	positioning,%
	shapes.geometric,%
	shapes,%
	snakes,%
}

\pgfplotsset{
    legend image with text/.style={
        legend image code/.code={%
            \node[anchor=center] at (0.3cm,0cm) {#1};
        }
    },
}

\newcommand{\be}{\begin{equation}}
\newcommand{\ee}{\end{equation}}
\newcommand{\bear}{\begin{eqnarray}}
\newcommand{\eear}{\end{eqnarray}}
\newcommand{\bears}{\begin{eqnarray*}}
\newcommand{\eears}{\end{eqnarray*}}
\newcommand{\bi}{\begin{itemize}}
\newcommand{\ei}{\end{itemize}}
\newcommand{\ben}{\begin{enumerate}}
\newcommand{\een}{\end{enumerate}}

\renewcommand{\ge}{\geqslant}

\newcommand{\R}{\mathbb{R}}
\newcommand{\E}{\mathbb{E}}

\definecolor{ogreen}{rgb}{0,0.5,0}
\definecolor{magenta}{rgb}{1.0, 0.11, 0.81}
\definecolor{mulberry}{rgb}{0.77, 0.29, 0.55}
\definecolor{xgray}{rgb}{0.9, 0.9, 0.9}
\def \blue{\color{blue}}

\definecolor{byzantium}{rgb}{0, 0, 0.5}

\def \bes{\begin{equation*}}
\def \ees{\end{equation*}}
\def \bas{\begin{align*}}
\def \eas{\end{align*}}
\def \be{\begin{equation}}
\def \ee{\end{equation}}
\def \bbm{\begin{bmatrix}}
\def \ebm{\end{bmatrix}}
\def \cC{\mathcal{C}}
\def \cf{\mathcal{S}}
\def \ck{\mathcal{R}}

\def \cd{\mathcal{D}}

\def \CO{\texttt{CO}}
\def \RO{\texttt{RO}}

\def \bes{\begin{equation*}}
\def \ees{\end{equation*}}
\def \bas{\begin{align*}}
\def \eas{\end{align*}}
\def \bbm{\begin{bmatrix}}
\def \ebm{\end{bmatrix}}
\def \cf{\mathcal{S}}
\def \cd{\mathcal{D}}

\def \cC{\mathcal{C}}

\def \w{{\fontfamily{cmtt}\selectfont \text{W}}}
\def \m{{\fontfamily{cmtt}\selectfont \text{M}}}

\newtheorem*{theorem*}{Theorem}
\newtheorem{theorem}{Theorem}
\newtheorem{lemma}[theorem]{Lemma}
\newtheorem{claim}[theorem]{Claim}
\newtheorem{example}{Example}

\newtheorem{corollary}[theorem]{Corollary}

\newcommand{\eq}[1]{\begin{align*}#1\end{align*}}

\def \CO{\fontfamily{euf}\selectfont \text{CO}}
\def \RO{\fontfamily{euf}\selectfont \text{RO}}

\IEEEoverridecommandlockouts

\begin{document}

\newlength\figureheight
\newlength\figurewidth

\title{Minimizing Latency for Secure Coded Computing Using Secret Sharing via Staircase Codes\vspace{-0.2cm}}
\author{
\IEEEauthorblockN{Rawad Bitar, Parimal Parag, and Salim El Rouayheb
\thanks{
R. Bitar and S. El Rouayheb are with the ECE department of Rutgers University. 
P. Parag is with the ECE department of the Indian Institute of Science.}%
\thanks{ 
Emails: rawad.bitar@rutgers.edu, parimal@iisc.ac.in, salim.elrouayheb@rutgers.edu. }%
\thanks{
Part of the work was presented at ISIT, 2017 \cite{BPR17}.}%
\thanks{
The work of the first and last authors was supported in parts by ARL Grant W911NF-17-1-0032.
The work of the second author was supported by the Science and Engineering Research Board (SERB) under Grant No.~DSTO-1677. 
Any opinions, findings, and conclusions or recommendations expressed in this material are those of the authors and do not necessarily reflect the view of 
SERB.
}
}
\vspace{-1cm}
}

\maketitle

\begin{abstract}
We consider the setting of a Master server, $\m$,  who possesses confidential data (e.g., personal, genomic or medical data) and wants to run intensive computations on it, as part of a machine learning algorithm for example. 
The Master wants to distribute these computations to untrusted workers  who have volunteered or are incentivized  to help with this task. 
However,  the data must be kept  private (in an information theoretic sense) and not revealed to the individual workers. Some of the workers may be stragglers, e.g., slow or busy, and will take a random time to finish the task assigned to them. 
We are interested in reducing the delays experienced by the Master. We focus on linear computations as an essential operation in many iterative algorithms such as principal component analysis, support vector machines and other gradient-descent based algorithms. 
A classical solution is to use a linear secret sharing scheme, such as Shamir's scheme, to divide the data into secret shares on which the workers can perform linear computations. However, classical codes can provide straggler mitigation assuming a worst-case scenario of a fixed number of stragglers. We propose a solution based on new secure codes, called Staircase codes, introduced previously by two of the authors. Staircase codes allow flexibility in the number of stragglers up to a given maximum, and universally achieve the information theoretic limit on the download cost by the Master, leading to latency reduction. Under the shifted exponential model, we find upper and lower bounds on the Master's mean waiting time. We derive the distribution of the Master's waiting time, and its mean, for systems with up to two stragglers. For systems with any number of stragglers, we derive an expression that can give the exact distribution, and the mean, of the waiting time of the Master. We show that Staircase codes always outperform classical secret sharing codes. For instance, for codes with rate $k/n=1/2$ Staircase codes can lead to up to $59\%$ reduction in delay compared to classical secret sharing codes. We validate our results with extensive implementation on Amazon EC2 clusters.
\end{abstract}

\section{Introduction}
We consider the setting of distributed computing in which a server $\m$, referred to as Master,  possesses  {\em confidential}  data and wants  to perform  intensive computations on it.  $\m$ wants to divide these computations into smaller computational tasks and distribute them to $n$ {\em untrusted} worker machines that can perform these smaller tasks in parallel. The workers then return their results to the Master, who can  process them to obtain the result of its original task.

In this paper, we are interested in applications in which the worker machines do not belong to the same system or cluster as the Master. Rather, the workers are online computing machines that can be  hired or can volunteer to help the Master in its computations, e.g., crowdsourcing platforms like the SETI@home \cite{setiah} and folding@home \cite{fah} projects. The additional constraint, which we worry about here, is that  the workers cannot be trusted with the sensitive data, which must remain hidden from them.  Privacy could be achieved using fully homomorphic encryption that allows computing over encrypted data. However, homomorphic encryption incurs high computation and storage overheads \cite{homomorphic2}, which may not be feasible in certain applications. 

We propose information theoretic security to achieve the privacy requirement. Information theoretic security is typically used to provide privacy with no constraints on the computational power of the adversary (compromised workers). Our main motivation for information theoretic security is the low complexity of the resulting schemes (compared to homomorphic encryption).  The assumption that we have to make here is a limit on the number of workers colluding against the Master.

We focus on  linear computations (matrix multiplication) since they form a building block of many iterative algorithms, such as principal component analysis, support vector machines and other gradient-descent based algorithms \cite{speeding,DCG16}. The workers introduce random delays due to the difference of their workloads or network congestion. This causes the Master to wait for the slowest workers, referred to as stragglers in the distributed computing community~\cite{dean2008mapreduce,DB13}. Our goal is to reduce the aggregate delay experienced by the Master.

Privacy can be achieved by encoding the data, with random keys, using linear secret sharing codes~\cite{AF10} as illustrated in Example~\ref{ex:intro}. However, these codes are not specifically designed to minimize latency as we will highlight later.

\begin{example}
\label{ex:intro}
\begin{figure}[h!]
\vspace{-0.8cm}
\centering
\begin{minipage}[t]{0.45\textwidth}
\centering
\resizebox{!}{0.45\textwidth}{ \begin{tikzpicture}[>=stealth', auto,
 triangle/.style = {fill=white, regular polygon, regular polygon sides=3 },]
  \definecolor{lightgray}{rgb}{0.83, 0.83, 0.83}
\def\mx{3}
\def\my{-4}
\tikzstyle{server} = [fill=black!10, rectangle, rounded corners=4mm, draw,minimum width=2em, minimum height=2.5em]

\node[inner sep=0] (s1) at (-4.3,0) {\includegraphics[height=9mm]{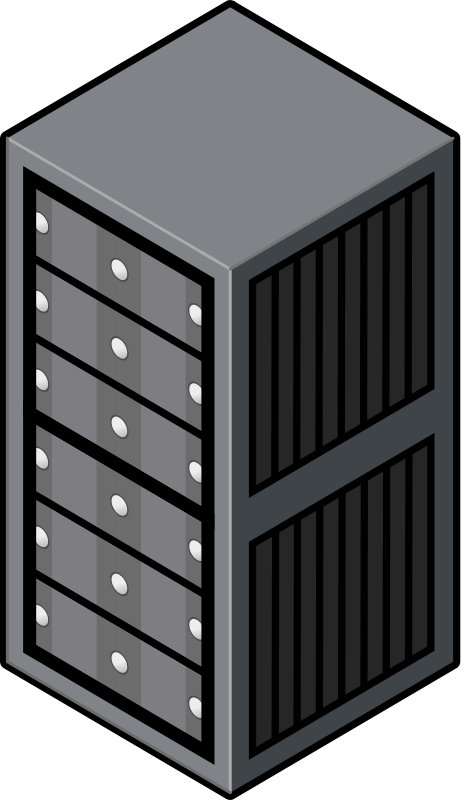}} node[below=0cm of s1,color=byzantium] {Master} node[left=0cm of s1,color=byzantium] {$\m$};

\node[server,right=0.8cm of s1,font=\footnotesize] (bb) {Encoding}; 
\node[above=0.8cm of bb, font=\footnotesize] (rm) {Randomness};

\node[inner sep=0pt,right =0.75 of bb,font=\footnotesize] (us2){\includegraphics[height=7mm]{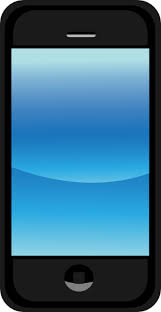}}
node[right =0 of us2,font=\small,color=byzantium] {$\w_2$};

\node[inner sep=0,above=0.35 cm of us2,font=\footnotesize] (us1) {\includegraphics[width=5mm]{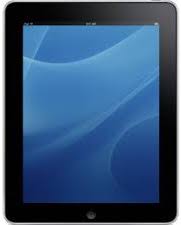}}
node[right= 0 cm of us1,font=\small,color=byzantium] {$\w_1$};

\node[inner sep=0,below= 0.35 cm of us2,font=\footnotesize] (us3) {\includegraphics[width=6mm]{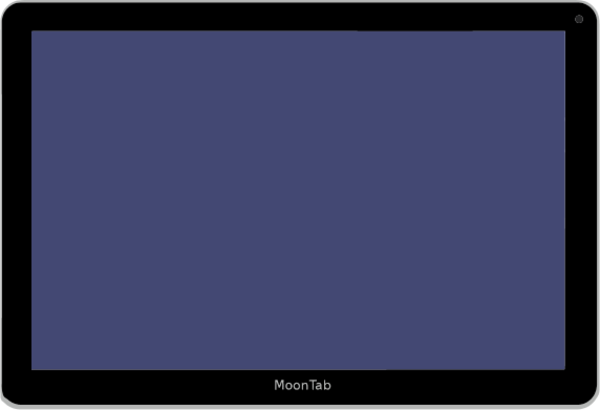}}
node[right=0 of us3,font=\small,color=byzantium] {$\w_3$} ;
 
\path[every node/.style={},color=black,->]
(s1) edge node[midway,below=0,font=\footnotesize] {$A$} node[midway,above=0,font=\footnotesize] {Data} (bb) 
   (bb.north east)    	edge  node[midway,above=0,sloped,font=\footnotesize] {$S_1$}  (us1) 
   (bb.east)		edge  node[midway,above=0,sloped,font=\footnotesize] {$S_2$}  (us2)
   (bb.south east)	edge  node[midway,below=0,sloped,font=\footnotesize] {$S_3$}  (us3)
   (rm) 			edge  node[midway,right=0,font=\footnotesize] {$R$} (bb);
   
\node[above=0 of us1,color=byzantium] {Workers};
\end{tikzpicture}
}
\captionsetup{subtype,width=\textwidth}
\caption{The Master $\m$ encodes its data $A$ with a random matrix $R$ into $3$ secret shares $S_1,\ S_2,\ S_3$. Any two shares can decode $A$. For example, $S_1= R$, $S_2= A+R$, and \mbox{$S_3=A+2R$}. $\m$ sends the share $S_i$ to worker $\w_i$. The randomness $R$ is used to ensure privacy.}
\label{fig:introfig1}
\end{minipage} \hfill%
\begin{minipage}[t]{0.45\textwidth}
\centering
\resizebox{!}{0.5\textwidth}{\begin{tikzpicture}
\node[inner sep=0] (s) at (0.7,0) {\includegraphics[height=9mm]{spymobile.png}} node[left=0cm of s,color=byzantium] {$\m$};

\node[inner sep=0pt,right =3.5 of s,font=\footnotesize] (us2){\includegraphics[height=7mm]{phone.jpg}}
node[left =-0.1 of us2,font=\small,color=byzantium] {} node[below =0 of us2,font=\footnotesize] {$S_2$} node[right =0 of us2,font=\small,color=byzantium] {$\w_2$};;

\node[inner sep=0,above= 0.6 cm of us2,font=\footnotesize] (us1) {\includegraphics[width=5mm]{tablet.jpg}}
node[left=-0.1cm of us1,font=\small,color=byzantium] {} node[below=0 of us1,font=\footnotesize] {$S_1$} node[right =0 of us1,font=\small,color=byzantium] {$\w_1$};;

\node[inner sep=0,below= 0.6 cm of us2,font=\footnotesize] (us3) {\includegraphics[width=6mm]{tablet2.png}}
node[left=-0.1 of us3,font=\small,color=byzantium] {} node[below= 0cm of us3,font=\footnotesize] {$S_3$} node[right =0 of us3,font=\small,color=byzantium] {$\w_3$};;

\path[every node/.style={},color=black,->]
   (s.north east)    	edge[bend left]  node[midway,above left=0 and -0.4 cm,sloped,font=\footnotesize] {$\mathbf{x}$} (us1) 
   (s.east)		edge[bend left]  node[midway,below=0,sloped,font=\footnotesize] {$\mathbf{x}$}  (us2)
   (s.south east)		edge[bend right]  node[midway,below=0,sloped,font=\footnotesize] {$\mathbf{x}$}   (us3)
			
   (us1)		edge[dashed]  node[midway,above =0, sloped,font=\footnotesize]  {$S_1\mathbf{x}$}  	(s.north east) 
   (us2)		edge[dashed]  node[midway,below =0, sloped,font=\footnotesize]  {$S_2\mathbf{x}$}  	(s)
   (us3)		edge[dashed]  node[midway,below =0, sloped,font=\footnotesize]  {$S_3\mathbf{x}$}  	(s.south east)
   ;
   
\end{tikzpicture}}
\captionsetup{subtype,width=\textwidth}
\caption{To compute $A\mathbf{x}$, $\m$ sends $\mathbf{x}$ to all the workers. Each worker $\w_i$ computes $S_i\mathbf{x}$ and sends the result to $\m$. $\m$ can decode $A\mathbf{x}$ after receiving any two responses, e.g., \mbox{$A\mathbf{x}=S_2\mathbf{x} - S_1\mathbf{x}= (A+R)\mathbf{x} - R\mathbf{x}$}.}
\label{fig:introfig2}
\end{minipage}
\caption{Secure distributed matrix multiplication with $3$ workers. The Master encodes its data using a linear secret sharing code, e.g., Shamir's codes (given in the caption) \cite{S79,McESa81} or Staircase codes (given in Table~\ref{tab:CESS}) \cite{BR16,BRIT17}. Decoding $A\mathbf{x}$ follows from the linearity of the code.}\label{fig:intro1}
\end{figure}
Let the matrix $A$ denote the data set owned by $\m$ and let $\mathbf{x}$ be a given vector. $\m$ wants to compute $A\mathbf{x}$. 
Suppose that $\m$ gets the help of $3$ workers out of which at most \mbox{$1$} may be a straggler.
$\m$ generates a random matrix $R$ of same dimensions as $A$ with entries drawn over the same alphabet as the entries of $A$. 
$\m$ encodes $A$ and $R$ into 3 shares $S_1=R$, $S_2=R+A$ and $S_3=R+2A$ using a secret sharing scheme~\cite{S79,McESa81}. 
$\m$ sends  share $S_i$ to worker $\w_i$ (Figure~\ref{fig:introfig1}) and then sends $\mathbf{x}$ to all the workers. 
Each worker computes $S_i\mathbf{x}$ and sends it back to $\m$ (Figure~\ref{fig:introfig2}). 
$\m$ can decode $A\mathbf{x}$ after receiving any $2$ responses. 
For instance, if the first two workers respond, $\m$ can obtain $A\mathbf{x}= S_2\mathbf{x}-S_1\mathbf{x}$. 
No information about $A$ is revealed to the workers, because $A$ is one-time padded by $R$.
\end{example}

In the previous example, even if there were no stragglers, $\m$ still has to wait for the full responses of two workers, and the response of the third one will not be used for decoding. In addition, $\m$ always has to decode $R\mathbf{x}$ in order to decode $A\mathbf{x}$. Hence, more delays are incurred by spending communication and computation resources on decoding $R\mathbf{x}$, which is only needed for privacy.
We overcome those limitations by using Staircase codes introduced in~\cite{BR16,BRIT17} which do not always require decoding $R\mathbf{x}$. Thus, possibly reducing the computation load at the workers and the communication cost at the Master. In addition, Staircase codes allow more flexibility in the number of responses needed for decoding $A\mathbf{x}$, as explained in the next example.

\begin{example}[Staircase codes]
\label{ex:intro2}
Consider the same setting as Example~\ref{ex:intro}. Instead of using a classical secret sharing code, $\m$ now encodes $A$ and $R$ using the Staircase code given in Table~\ref{tab:CESS}.
\begin{table}[t] 
\centering
\begin{tabular}[h!]{c|c|c}
$S_1$ & $S_2$ & $S_3$ \\ \hline
$A_1+A_2+R_1$ & $A_1+2A_2+4R_1$ &$A_1+3A_2+4R_1$\\
\blue $R_1+R_2$ & \blue $R_1+2R_2$ & \blue $R_1+3R_2$ \\
\end{tabular}
 \caption{The shares sent by $\m$ to each worker when using Staircase codes. In this example, each share is divided into two sub-shares giving the Master more decoding options. The Master can decode $A\mathbf{x}$ by receiving the multiplication of $\mathbf{x}$ by either the first sub-share of each worker (in black) or two sub-shares from any two workers (in black and blue). Note that if $\m$ uses the first three sub-shares, it only decodes half of $R\mathbf{x}$, i.e., $R_1\mathbf{x}$, and does not need to decode $R_2\mathbf{x}$. The operations shown are in $GF(5)$.}
 \label{tab:CESS}
 \vspace{-0.2cm}
\end{table}
The Staircase code requires $\m$ to divide the matrices $A$ and $R$ into $A=\begin{bmatrix}A_1& A_2\end{bmatrix}^T$ and $R=\begin{bmatrix}R_1& R_2\end{bmatrix}^T$. In this setting, $\m$ sends two sub-shares to each worker, hence each task consists of $2$ sub-tasks. The Master sends $\mathbf{x}$ to all the workers. Each worker  multiplies the sub-shares by $\mathbf{x}$ (going from top to bottom) and sends each multiplication back to $\m$ independently. Now, $\m$ has two possibilities for decoding:
  \begin{enumerate*}[label=\emph{\arabic*)}] \item $\m$ receives the first sub-task from all the workers, i.e., receives $(A_1+A_2+R_1)\mathbf{x}$, $(A_1+2A_2+4R_1)\mathbf{x}$ and $(A_1+2A_2+4R_1)\mathbf{x}$ and decodes $A\mathbf{x}$ which is the concatenation of $A_1\mathbf{x}$ and $A_2\mathbf{x}$. Note that here $\m$ decodes only $R_1\mathbf{x}$ and does not need to decode $R_2\mathbf{x}$. \item $\m$ receives all the sub-tasks from any $2$ workers and decodes $A\mathbf{x}$. Here $\m$ has to decode $R_1\mathbf{x}$ and $R_2\mathbf{x}$.
 \end{enumerate*}
One can check that no information about $A$ is revealed to the workers, because each sub-share is padded by a random matrix.\end{example}

Under a shifted exponential delay model for each worker, we show that the Staircase code given in Example~\ref{ex:intro2} can lead to a $45\%$ improvement in delay over the secret sharing code given in Example~\ref{ex:intro}. Our goal is to give a general systematic study of the delay incurred by Staircase codes and compare it to classical secret sharing codes.

\subsection{Contributions} 
To the extent of our knowledge, this paper is the first work to study privacy requirement for distributed coded computing. We consider the distributed computing setting described above in which we require the workers to learn no information (in an information theoretic sense) about the Master's data. We study the waiting time of the Master caused by delays of the workers. We follow the literature, e.g., \cite{TOFEC, speeding}, and model the service time at the workers as a shifted exponential random variable. This service time includes upload time, computation time and download time. Finding codes that minimize the delay at the Master is still an open problem in general. In this work, we take the download communication cost as a proxy for delay when designing the coding schemes. More precisely, we study the performance of the recently introduced Staircase codes \cite{BR16,BRIT17} that achieve the information theoretic limit on download cost \cite{HLKBtrans} and compare them to classical secret sharing codes.

Before we state our contributions, we introduce some necessary notations. We denote by $n$  the number of workers available to help the Master, $k$ denote the minimum number of non stragglers and $z$  the maximum number of colluding workers. We refer to such secure distributed computing system by an $(n,k,z)$ system. 
We make the following contributions:%
\begin{enumerate}
\item  {\em General bounds for systems with any number of stragglers:} We derive an upper and a lower bound on the Master's mean waiting time when using Staircase codes (Theorem~\ref{thm:ublb}). Moreover, we derive the exact distribution of the Master's waiting time when using Staircase codes, in an integral form (Theorem~\ref{thm:exactexp}).  Using the upper bound, we compare the performance of Staircase codes to classical secret sharing codes and characterize the savings obtained by Staircase codes. We show that Staircase codes always outperform classical secret sharing codes.

\item {\em Exact characterization for systems with up to $2$ stragglers:}  We use the integral expression of Theorem~\ref{thm:exactexp} to find the exact distribution of the Master's waiting time for systems with up to $n-k=1$ and up to $n-k=2$ stragglers (Corollary~\ref{Corr:ExactWaiting}). Moreover,  we derive the exact expressions of the Master's mean waiting time for these systems (Theorem~\ref{Corr:MeanWaiting}) and use these expressions to show the tightness of our upper bound.

\item {\em Simulations and validation:} We ran extensive MATLAB simulations for different system parameters. We focus on two regimes: regime with fixed rate $k/n$ and regime with fixed maximum number of stragglers $n-k$. Our main observation is that the upper bound, based on Jensen's inequality, is a good approximation of the mean waiting time.  Furthermore, we validate our results with extensive implementation on Amazon EC2 clusters. The savings obtained on EC2 clusters are within the range of the values predicted by the theoretical model. To give an example, for $n=4$ workers, large data and high traffic regime, our implementation shows $59\%$ (Figure~\ref{fig:amzwaiting42}) savings in the mean waiting time while the theoretical model predicts $66\%$ savings (Figure~\ref{fig:savings1}). 
\end{enumerate}

\subsection{Related work} 

The problem of stragglers has been identified and studied by the distributed computing community, see e.g., \cite{dean2008mapreduce,zaharia2008improving,DB13,chen2016revisiting,AKGSLSH10,narayanamurthy2013towards,recht2011hogwild,mitliagkas2016asynchrony,RRTB12, ZCJL13, YYH+13, LWR+14, DJM13, WSD+14, HYD15, mania2015perturbed,chilimbi2014project,dean2012large}. %
Recently, there has been a growing research interest in studying codes for straggler mitigation and delay minimization in distributed systems with no secrecy constraints. The early body of work focused on minimizing latency of content download in distributed storage systems, see e.g., \cite{HPZR12,JLS12,TOFEC,KSS15} and later the focus has shifted to using codes for straggler mitigation in distributed computing.

In~\cite{speeding} Lee et al.\ studied the use of MDS codes for straggler mitigation in {\em linear} distributed machine learning algorithms. 
Yu et al. \cite{yu2017polynomial} introduced a coding scheme called polynomial codes to mitigate straggler in distributed matrix multiplication.
 Tandon et al.\ \cite{tandon2016gradient} introduced a framework called gradient coding for straggler mitigation in distributed gradient-descent based algorithms. In the same spirit of work, Halbawi et al.~\cite{halbawi2017improving} proposed a gradient coding scheme that  decreases the decoding complexity at the Master.

In~\cite{DCG16}, Dutta et al.\ proposed new coding techniques that reduce the computation time at the workers side while accounting for stragglers. Moreover, coded computing was studied for specific applications, such as coded convolution~\cite{dutta2017coded} and coded linear transformations~\cite{yang2017computing}. 
In a related context, Li et al.~\cite{li2016unified,li2016fundamental} showed a fundamental tradeoff between the workers' computation load and the communication complexity in coded computing.

The work in the literature on coded computing has not  considered so far privacy constraints. In general, privacy in distributed computing is studied separately, mostly in the computer science community. The work that is closest to ours is~\cite{AF10} that studies the problem of distributively multiplying two private matrices under information theoretic privacy constraints using classical secret sharing codes.  
Our work can also be related to the work on privacy-preserving algorithms, e.g.,~\cite{secureml1,secureml2,secureml3,gade2017private}. 
However, the privacy constraint in this line of work is computational privacy, and the proposed algorithms are not designed for straggler mitigation.

 \subsection{Organization} 
 The paper is organized as follows. 
 We formalize the problem and define the model in section~\ref{sec:sysmod}. 
 In section~\ref{sec:main}, we present and discuss our main results. We describe the construction of Staircase codes and classical secret sharing codes in Section~\ref{sec:encoding}. 
In Sections~\ref{sec:proof1} and~\ref{sec:proof2}, we study the probability distribution of the Master's waiting time and derive bounds on the mean waiting time. 
We show, in Section~\ref{sec:imp}, that the (random) number of workers that minimizes the waiting time is concentrated around its average. We evaluate the performance of Staircase codes via simulation in Section~\ref{sec:simu}. In Section~\ref{sec:AWS}, we give a representative sample of our implementation on Amazon EC2 clusters and compare them to our theoretical findings. We conclude the paper in section~\ref{sec:conc}. We prove Theorem~\ref{Corr:MeanWaiting} and discuss how our scheme can be extended to provide secrecy over multiple iterations of an algorithm.

\section{System Model} 
\label{sec:sysmod}

We consider a Master server $\m$ which wants to perform intensive computations on confidential data represented by an $m \times \ell$ matrix $A$ (typically $m>>\ell$). In machine learning applications $m$ denotes the number of data points (examples) possessed by $\m$ and $\ell$ denotes the number of attributes (features) of each example.
$\m$ divides these computations into smaller computational tasks and assigns them to $n$ workers $\w_i$, $i=1,\dots,n$, that can perform these tasks in parallel. The division is horizontal, i.e., each worker gets a given number of rows of $A$ with all their corresponding columns.

\subsection{Computations model} 
We focus on linear computations. The motivation is that  a building block in several iterative machine learning algorithms, such as gradient descent, is the multiplication of $A$ by a  sequence of  $\ell \times 1$ attribute vectors $\mathbf{x}^1, \mathbf{x}^2, \dots$. 
In the sequel, we focus on the multiplication $A\mathbf{x}$ with one attribute vector $\mathbf{x}$.

\subsection{Workers model} 
The workers have the following properties: 
\begin{enumerate*}[label=\emph{\arabic*)}]
\item {The workers incur random delays while executing the task assigned to them by $\m$ resulting in what is known as the straggler problem~\cite{dean2008mapreduce,DB13,speeding}.} 
We model all the delays incurred by each worker by an independent and identical shifted exponential random variable. \item Up to $z$, $z<k$, workers can collude, i.e., at most $z$ workers can share with each other the data they receive from $\m$. The threshold $z$ could be thought of as a desired level of security. This has implications on the privacy constraint described later.\end{enumerate*}

\subsection{General scheme} 
$\m$ encodes $A$, using randomness, into $n$ shares $S_i$ sent to worker $\w_i$, $i=1,\dots,n$. Any $k$ or more shares can decode $A$, and any collection of $z$ workers obtain zero information about $A$. 
For any set $\mathcal{B}\subseteq \{1,\dots,n\}$, let $S_\mathcal{B}=\{S_i, i\in\mathcal{B}\}$ denote the collection of shares given to worker $\w_i$ for all $i\in \mathcal{B}$. %
The previous requirements can be expressed as,
\begin{align*}
H(A|S_{\mathcal{B}})&=0,\quad \forall \mathcal{B}\subseteq \{1,\dots,n\} \text{ s.t. } |\mathcal{B}|\geq k,\\
 H(A|S_\mathcal{Z})&=H(A),\quad  \forall \mathcal{Z}\subseteq \{1,\dots,n\} \text{ s.t. } |\mathcal{Z}|\leq z.
 \end{align*}

\noindent At each iteration, the Master sends $\mathbf{x}$ to all the workers. Then, each worker computes $S_i\mathbf{x}$ and sends it back to the Master. Since the scheme and the computations are linear, the Master can decode $A\mathbf{x}$ after receiving enough responses. We refer to such scheme as an $(n,k,z)$ system. In some cases the attribute vectors $\mathbf{x}$ contain information about $A$, and therefore need to be hidden from the workers. We describe in Section~\ref{sec:hiding} how our scheme can be generalized to such cases.

\subsection{Encoding} 
We consider classical secret sharing codes~\cite{S79,McESa81} and universal Staircase codes\cite{BR16,BRIT17}. 
We now describe their properties that are necessary for the delay analysis, 
with more detailed descriptions in Section~\ref{sec:encoding}. 
Secret sharing codes require the division of $A$ into $k-z$ row blocks and encodes them into $n$ shares of dimension $\frac{m}{(k-z)}\times \ell$ each. 
Any $k$ shares can decode $A$. 
Similarly, Staircase codes encode $A$ into $n$ shares of $\frac{m}{(k-z)}\times \ell$ each with the additional requirement that each share is divided into  $b=\text{LCM}\{k-z+1,\dots,n-z\}$ sub-shares. 
Decoding $A$ requires a fraction $\alpha_d b$ sub-shares, $\alpha_d \triangleq \frac{(k-z)}{(d-z)},$ from any of the $d$ shares, $d\in \{k,\dots,n\}$. 
We show that Staircase codes outperform classical codes in terms of incurred delays.

\subsection{Delay model} 
Let $T_A$ be the random variable representing the time spent to compute $A\mathbf{x}$ at one worker. 
We assume a mother runtime distribution $F_{T_A}(t)$ that is shifted exponential with rate $\lambda$ and a constant shift $c$. 
For each $i \in\{1,\dots,n\}$, we let $T_i$ denote the time spent by worker $\w_i$ to execute its task. 
Due to the encoding, each task given to a worker is $k-z$ times smaller than $A$, 
or $T_i = \frac{T_A}{(k-z)}$. 
It follows that $F_{T_i}$ is a scaled distribution of $F_{T_A}$. 
That is, 
\begin{equation}
F_{T_i}(t)\triangleq F_{T_A}((k-z)t)=1-e^{ -\lambda (k-z)(t-\tfrac{c}{k-z})}, \quad \text{for } t\geq c/(k-z).
\end{equation}
For an $(n,k,z)$ system using Staircase codes, we assume that $T_i$ is evenly distributed among the sub-tasks\footnote{Therefore, we make  two assumptions on the waiting time of the sub-tasks: (1) the parameters of its distribution (effective $c$ and $\lambda$) vary linearly with the sub-task size and (2) the waiting time of  sub-tasks of the same task take equal service time, and therefore are not independent. These assumptions make the problem more amenable to theoretical analysis. In Section~\ref{sec:AWS}, we compare our model to traces obtained from Amazon cloud and show that our model provides insightful engineering guidelines.}. That is, the time spent by a worker $\w_i$ on one sub-task is equal to $T_i/b$, 
and the time spent on $b\alpha_d = b\frac{k-z}{d-z}$ sub-tasks is $\alpha_dT_i$.  

Let $T_{(i)}$ be the $i^{th}$ order statistic of the $T_i$'s and $T_{\text{SC}}(n,k,z)$ be the time the Master waits until it can decode $A\mathbf{x}$. 
If the aggregate wait is due to $d$ workers each finishing $\alpha_d$ fraction of its $b$ sub-tasks, 
then the Master's waiting time is $\alpha_dT_{(d)}$. 
We can write
\begin{equation}\label{eq:sub}
T_{\text{SC}}(n,k,z)=\min_{d\in\{k,\dots,n\}}\left\{\alpha_dT_{(d)}\right\}. %
\end{equation}
It is useful for our analysis to look at $T_i$ as the sum of an exponential random variable $T'_i$ and a constant offset, i.e.
\begin{equation*}
T_{i}=T'_{i}+c/(k-z), \text{ where } T'_{i} \sim \exp{(\lambda(k-z))}.
\end{equation*}
From this interpretation, it is easy to verify that the $d^{\text{th}}$ order statistic $T_{(d)}$ of $(T_1, T_2, \dots, T_n)$ can be expressed as
\begin{equation*}
T_{(d)}=T'_{(d)}+c/(k-z),
\end{equation*}
where $T'_{(d)}$ is the $d^{\text{th}}$ order statistic of $n$ \emph{iid} exponential random variables with rate $\lambda(k-z)$. Therefore, we can write the Master's waiting time for Staircase codes as
\begin{equation}\label{eq:tsc}
T_{\text{SC}}(n,k,z) = \min_{d\in\{k,\dots,n\}} \left\{\alpha_d\Big(T'_{(d)}+\frac{c}{k-z}\Big)\right\}.
\end{equation}
For an $(n,k,z)$ system using classical secret sharing codes, 
the Master's waiting time $T_{\text{SS}}(n,k,z)$ is equal to the time spent by the fastest $k$ workers to finish their individual tasks. Hence, we can write 
\begin{equation}\label{eq:tss}
T_{\text{SS}}(n,k,z)=T_{(k)}.
\end{equation}
We drop the $(n,k,z)$ notation from $T_{\text{SC}}(n,k,z)$ and $T_{\text{SS}}(n,k,z)$ when the system parameters are clear from the context.

\begin{figure}[t]
\centering
 \setlength\figureheight{0.35\textwidth}
  \setlength\figurewidth{0.4\textwidth}
\begin{minipage}[b]{0.32\textwidth}
  \centering
\captionsetup{subtype,width=\textwidth}
\resizebox{0.9\textwidth}{!}{
\definecolor{mycolor1}{rgb}{0.0, 0.58, 0.71}
\definecolor{mycolor3}{rgb}{1.0, 0.65, 0.0}%
\definecolor{mycolor2}{rgb}{0.8, 0.31, 0.36}
\def \msize{5}
\begin{tikzpicture}

\begin{axis}[%
width=0.951\figurewidth,
height=\figureheight,
at={(0\figurewidth,0\figureheight)},
scale only axis,
unbounded coords=jump,
xmin=0,
xmax=60,
xlabel={Number of workers $n$},
xlabel style = {font=\Large},
ymin=0,
ymax=1.1,
ylabel={Mean waiting time},
ylabel style = {font=\Large},
axis background/.style={fill=white},
title style={font=\bfseries},
legend style={legend cell align=left,align=left,draw=white!15!black}
]
\addplot [color=mycolor1,mark=o,mark options={solid},dashed,ultra thick]
  table[row sep=crcr]{%
4	1.02777777777778\\
6	0.6125\\
8	0.443571428571429\\
10	0.346995464852608\\
15	0.234852416102416\\
20	0.182219979831981\\
25	0.150725371716069\\
40	0.102122784836118\\
50	0.0850894752836048\\
60	0.0733310598763463\\
80	0.0579932373889028\\
100	0.048323479563295\\
};
\addlegendentry{Upper bound in \eqref{eq:main2}};

\addplot [color=mycolor2, only marks, mark=+,mark options={solid}, ultra thick,mark size = \msize]
  table[row sep=crcr]{%
4	0.899510330967132\\
6	0.553829406983357\\
8	0.410215730891787\\
10	0.331058889619674\\
15	0.230236389804481\\
20	0.180209143290515\\
25	0.149652043837141\\
40	0.101828210716796\\
50	0.0849286620348703\\
60	0.073232750521484\\
80	0.0579478882678339\\
100	0.0482985630301599\\
};
\addlegendentry{Mean waiting time in \eqref{eq:corr12}};

\addplot [color=mycolor3,solid,ultra thick,mark=o,mark options={solid},]
  table[row sep=crcr]{%
4	0.571428571428571\\
6	0.405688429217841\\
8	0.323493919608738\\
10	0.272765767418202\\
15	0.201302070944928\\
20	0.163038929323352\\
25	0.138164924073063\\
40	0.0968857189470866\\
50	0.0816164354761107\\
60	0.0708452612364701\\
80	0.0565250541638673\\
100	0.0473472476529255\\
};
\addlegendentry{Lower bound in \eqref{eq:main1}};

\end{axis}
\end{tikzpicture}
}
\caption{Systems with fixed $n-k=2$.}
\label{fig:thm1}
\end{minipage}\hfill %
\begin{minipage}[b]{0.32\textwidth}
\centering
\captionsetup{subtype,width=\textwidth}
\resizebox{0.85\textwidth}{!}{
\definecolor{mycolor1}{rgb}{0.0, 0.58, 0.71}
\definecolor{mycolor3}{rgb}{1.0, 0.65, 0.0}
\definecolor{mycolor4}{rgb}{0.53, 0.66, 0.42} 
\definecolor{mycolor2}{rgb}{0.8, 0.31, 0.36}
\begin{tikzpicture}
\def \msize{5}
\begin{axis}[%
width=0.951\figurewidth,
height=\figureheight,
at={(0\figurewidth,0\figureheight)},
scale only axis,
unbounded coords=jump,
xmin=0,
xmax=80,
xlabel={Number of workers $n$},
xlabel style = {font=\Large},
ymin=0,
ymax=0.45,
ylabel={Mean waiting time},
ylabel style = {font=\Large},
axis background/.style={fill=white},
title style={font=\bfseries},
legend style={legend cell align=left,align=left,draw=white!15!black}
]
\addplot [color=mycolor1,dashed, ultra thick,mark=o,mark options={solid}]
  table[row sep=crcr]{%
8	0.309821428571429\\
10	0.244926303854875\\
14	0.170525560298288\\
20	0.117146877460345\\
24	0.0970173567894493\\
40	0.0574304176786973\\
50	0.0457722703123227\\
60	0.0380406158458476\\
80	0.0284348931960061\\
100	0.0227049627754698\\
};
\addlegendentry{Upper bound in \eqref{eq:main2}};

\addplot [color=mycolor2,only marks,mark=+,mark options={solid}, ultra thick,mark size = \msize]
  table[row sep=crcr]{%
8	0.288275039956867\\
10	0.226526212865715\\
14	0.15913057446545\\
20	0.110479141858043\\
24	0.0918266347510298\\
40	0.0550447845403293\\
50	0.0440766402838816\\
60	0.0367416561866295\\
80	0.0276153581002371\\
100	0.0221124153690991\\
};
\addlegendentry{Staircase codes};

\addplot [color=mycolor4,solid,ultra thick,mark=o,mark options={solid},mark size = 1]
  table[row sep=crcr]{%
8	0.439689314289829\\
10	0.330513610594016\\
14	0.221539835208388\\
20	0.148286483315256\\
24	0.121248662961719\\
40	0.0705100952304872\\
50	0.055897970354654\\
60	0.0462843053452394\\
80	0.0344576198018599\\
100	0.0274289425162648\\
};
\addlegendentry{Classical codes};

\addplot [color=mycolor3,solid,ultra thick,mark=o,mark options={solid}]
  table[row sep=crcr]{%
8	0.193615645880028\\
10	0.150431553844602\\
14	0.104001966280689\\
20	0.0710788325029196\\
24	0.0586902835569462\\
40	0.0345792974657015\\
50	0.0275142313548418\\
60	0.0228462944445485\\
80	0.0170581800894308\\
100	0.0136100377272795\\
};
\addlegendentry{Lower bound in \eqref{eq:main1}};

\end{axis}
\end{tikzpicture}
}\vspace{0.2cm}
\caption{Systems with fixed rate $k/n=1/2$.}
\label{fig:thm2}
\end{minipage}\hfill %
\begin{minipage}[b]{0.32\textwidth}
  \centering
\captionsetup{subtype,width=\textwidth}
\resizebox{0.9\textwidth}{!}{
\definecolor{mycolor1}{rgb}{0.0, 0.58, 0.71}
\definecolor{mycolor3}{rgb}{1.0, 0.65, 0.0}
\definecolor{mycolor4}{rgb}{0.8, 0.31, 0.36}
\begin{tikzpicture}

\begin{axis}[%
width=0.951\figurewidth,
height=\figureheight,
at={(0\figurewidth,0\figureheight)},
scale only axis,
xmin=0,
xmax=100,
xlabel={Number of workers $n$},
xlabel style = {font=\Large},
ymin=0,
ymax=0.9,
ytick={0,0.1,0.2,0.3,0.4,0.5,0.6,0.7,0.8,0.9,1,1.1,1.2,1.3,1.4},
yticklabels={{  0\%},{ 10\%},{ 20\%},{ 30\%},{ 40\%},{ 50\%},{ 60\%},{ 70\%},{ 80\%},{ 90\%},{100\%},{110\%},{120\%},{130\%},{140\%}},
ylabel={Percentage of time savings},
ylabel style = {font=\Large},
axis background/.style={fill=white},
title style={font=\bfseries,align=center},
legend style={legend cell align=left,align=left,draw=white!15!black}
]

\addplot [color=mycolor3,ultra thick,,solid,mark=o,mark options={solid},mark size = 3]
  table[row sep=crcr]{%
4	0.348145892254733\\
8	0.188436999334596\\
10	0.154958099785173\\
14	0.127120725309053\\
20	0.107924350810415\\
24	0.101887519201008\\
40	0.0906403131712038\\
50	0.0864985774447925\\
60	0.0843264409490806\\
80	0.0794424924953238\\
100	0.0777918867060859\\
};
\addlegendentry{$k/n=1/2$};

\addplot [color=mycolor1,ultra thick,solid,mark=square,mark options={solid},mark size = 3]
  table[row sep=crcr]{%
8	0.650182208902922\\
16	0.509535269714274\\
24	0.467358519234207\\
40	0.435259809093379\\
48	0.427418600805132\\
60	0.419114621150682\\
80	0.411982584154399\\
100	0.407248872034928\\
};
\addlegendentry{$k/n=1/4$};

\addplot [color=mycolor4,ultra thick,solid,mark=triangle,mark options={solid},mark size = 3]
  table[row sep=crcr]{%
10	0.713201497227346\\
20	0.593462217036305\\
30	0.555445435474639\\
40	0.538260518432543\\
50	0.527263989127932\\
60	0.519463474032016\\
80	0.511101149529576\\
100	0.506036016103199\\
};
\addlegendentry{$k/n=1/5$};

\end{axis}
\end{tikzpicture}
}
\caption{Savings for systems with fixed $k/n$.}
\label{fig:thm3}
\end{minipage}
\caption{Theoretical upper and lower bounds for systems with rate of the exponential random variable $\lambda =1$, shift $c = 1$ and no colluding workers, i.e., $z=1$. Figure~\ref{fig:thm1} compares the bounds derived in Theorem~\ref{thm:ublb} to the theoretical mean waiting time for $(k+2,k,1)$ derived in Corollary~\ref{Corr:MeanWaiting}. Observe that the upper bound in \eqref{eq:main1} is a good approximation of the mean waiting time in \eqref{eq:corr12}. Figure~\ref{fig:thm2} compares the bounds in \eqref{eq:main1} and \eqref{eq:main2} to the simulated mean waiting time for $(n,k,z)$ systems with fixed rate $k/n=1/2$. We obtain the mean waiting time by averaging over $10000$ iterations for each value of $n$. Figure~\ref{fig:thm3} compares the upper bound in \eqref{eq:main1} to the mean waiting time of classical secret sharing in \eqref{eq:etss}. The savings are computed as the normalized difference between the waiting time of Staircase codes and classical secret sharing codes, i.e., \mbox{$\left(\E {[T_\text{SC}] }-\E {[T_\text{SS}] }\right)/\E {[T_\text{SS}] }$}.}
\label{fig:thm}
\end{figure}

\section{Our Results} 
\label{sec:main}
Our results characterize the delay performance of secure coded computing when using Staircase codes and compare it to classical secret sharing codes. The performance of Staircase codes is reflected in the Master's waiting time $T_{\text{SC}}$. 
Towards our goal, we establish in Theorem~\ref{thm:ublb} general bounds on the Master's mean waiting time $\E[T_\text{SC}(n,k,z)]$ when using Staircase codes for all $(n,k,z)$ systems, under the shifted exponential delay model. 
\begin{theorem}[Bounds on the Master's mean waiting time $\E {[T_\text{SC}] }$]
\label{thm:ublb}
Let $H_n$ be the $n^{\text{th}}$ harmonic sum defined as $H_n \triangleq \sum_{i=1}^n \frac{1}{i}$, 
with the notation $H_0 \triangleq 0$. 
The mean waiting time of the Master $\mathbb{E}[T_{\text{SC}}]$ for an $(n,k,z)$ Staircase coded system is upper bounded by
\begin{align}
\label{eq:main1}
\E[T_{\text{SC}}] &\leq \min_{d\in\{k,\dots,n\}}\left(\frac{H_n - H_{n-d}}{\lambda(d-z)}+\dfrac{c}{d-z}\right),
\end{align}
and lower bounded by 
\begin{align}
\label{eq:main2}
\E[T_{\text{SC}}] &\geq \frac{c}{n-z}+  \max_{d\in \{k,\dots,n\}}
\sum_{i=0}^{k-1}\binom{n}{i}\sum_{j=0}^{i}\binom{i}{j}\frac{2(-1)^j}{\lambda\left(2(n-i+j)(d-z)+(n-d)(n-d+1)\right)}.
\end{align}
\end{theorem}

We derive in Section~\ref{sec:proof2} a general integral expression \eqref{eq:exactexp} leading to the CDF $F_{T_\text{SC}}(t)$ of $T_\text{SC}$, the waiting time of the Master for all $(n,k,z)$ systems. Using the general integral expression, we derive the exact expression of the CDF  $F_{T_\text{SC}}(t)$  for systems with $n=k+1$ and $n=k+2$ as stated in the next Theorem. 

\newcounter{temp}
\begin{theorem}[Exact expression of $\E {[T_\text{SC}]} $ for systems with up to $2$ stragglers]
\label{Corr:MeanWaiting} 
The mean waiting time of the Master for $(k+1,k,z)$ and $(k+2,k,z)$ systems is given in \eqref{eq:corr11} and~\eqref{eq:corr12}, respectively.
\begin{align}
\label{eq:corr11}
\mathbb{E}\left[T_{\text{SC}}(k+1,k,z)\right] &= \frac{c}{k-z+1}+\frac{1}{\lambda}\sum_{i=1}^{k+1}(-1)^{i}\binom{k+1}{i}\left[\frac{i\exp\left(\frac{-\lambda c}{k-z}\right)}{(k-z)i+1}-\dfrac{1}{(k-z+1) i} \right].\\
\label{eq:corr12}
\E[T_{\text{SC}}(k+2,k,z)]&= \E[T_\text{SC}(k+2,k+1,z)]+\frac{1}{\lambda}\sum_{i=2}^{k+2}(-1)^i\binom{k+2}{i}\binom{i}{2}\left[\dfrac{\exp\left(-\frac{4 \lambda c}{k-z}\right)}{(k-z)i+4}-\dfrac{2\exp\left(-\frac{3 \lambda c}{k-z}\right)}{(k-z)i+3}\right].
\end{align}
\end{theorem}

To give insights into the theoretical bounds above, we compare in Figure~\ref{fig:thm1} bounds \eqref{eq:main1} and \eqref{eq:main2} for the case of $n=k+2$ to the exact expression in \eqref{eq:corr12}. We see that the upper bound in \eqref{eq:main1} is closer to the actual value and the gap between the two bounds closes as $n$ increases. We also establish the comparison for fixed rate regimes, in particular rate $k/n=1/2$. Since here $n\geq k+2$, we compare in Figure~\ref{fig:thm2} the bounds to numerical results obtained by simulation and observe the same behavior as before. We also plot in the same figure the mean waiting time for classical secret sharing codes obtained from \eqref{eq:tss} and given by
\begin{equation}\label{eq:etss}
\E[T_{\text{SS}}] = \frac{H_n - H_{n-k}}{\lambda(k-z)}+\dfrac{c}{k-z}.
\end{equation} 
This allows to verify that Staircase codes always outperform classical secret sharing codes. In Figure~\ref{fig:thm3}, we 
plot the lower bound on the relative savings brought by Staircase codes for systems with rate $k/n=1/2,1/4,1/5$. For instance, for rate $1/4$, the savings are lower bounded by $40\%$ for large $n$.  We supplement our theoretical results in Section~\ref{sec:simu} with an extensive array of simulations  in addition to measurement results obtained by implementation on Amazon EC2 clusters. The savings obtained in the implementation on Amazon cloud are within the savings predicted by the theoretical model.

\section{Staircase codes} \label{sec:encoding}
Staircase codes are the main ingredient of our scheme. The goal of this section is to explain the encoding and decoding of Staircase codes that are necessary for our delay analysis. Before we explain Staircase codes, we start by briefly explaining the encoding and decoding of classical secret sharing codes, which can be seen as a special case of Staircase codes. The main advantage of Staircase codes is allowing the Master to decode its data ($A\mathbf{x}$) without having to decode all the keys introduced to ensure privacy. This reduces both the computation load at the workers and the download cost at the Master, hence reduces the delays.

\subsection{Classical secret sharing}
Let $A$ be an $m \times \ell$ matrix with elements drawn uniformly at random from a finite alphabet, e.g., a finite field. An $(n,k,z)$ classical secret sharing (a.k.a. threshold secret sharing) code~\cite{S79,McESa81} allows the Master to encode the data $A$ into $n$ shares and distribute them to $n$ workers, such that any set of $z$, $z<k<n$, workers obtain no information about $A$. In addition, the Master can reconstruct $A$ by contacting any set of at least $k$, $k<n$, workers and downloading their shares. The construction requires the Master to divide $A$ into $k-z$ matrices $A_1,\dots,A_{k-z}$ of size $m/(k-z)\times \ell$ each. To ensure secrecy, threshold secret sharing codes require the use of $z$ matrices $R_1,\dots,R_z$ of dimensions $m/(k-z) \times \ell$. The elements of $R_i,\ i=1,\dots,z$, are drawn uniformly at random, and independently from $A$, from the finite alphabet to which the elements of $A$ belong.

\noindent{\em Encoding:} Let $GF(q),\ q\geq n$ denote the alphabet to which the elements of $A$ and $R$ belong. The encoding of an $(n,k,z)$ threshold secret sharing code~\cite{McESa81} can be seen as the multiplication of an $n \times k$ Vandermonde matrix\footnote{The necessary property of the encoding matrix $V$ is that any $k\times k$ square sub-matrix of $V$ is invertible. Vandermonde and Cauchy matrices satisfy this property.} $V$ defined over $GF(q)$ by the matrix $M_\text{SS}$ composed of the vertical concatenation of $A$ and $R$. The $i^{\text{th}}$ codeword, i.e., $i^{\text{th}}$ share  of the code, is the $i^{\text{th}}$ row of the matrix $C=VM_\text{SS}$.

\noindent{\em Decoding:} The decoding of threshold secret sharing consists of taking any $k$ codewords and inverting the corresponding encoding sub-matrix to obtain the secret and all the random keys. 

In the setting of distributed computing, the Master encodes $A$ and send the $n$ shares to the workers. To compute $A\mathbf{x}$, $\m$ sends $\mathbf{x}$ to the workers and waits for the first $k$ workers to send the multiplication of their shares by $\mathbf{x}$. Since the multiplication is linear, $\m$ can decode $A\mathbf{x}$ and $R\mathbf{x}$.

The following example illustrates the encoding and decoding of classical secret sharing.

\begin{example}[Classical secret sharing]
\label{ex:SS}
Consider the same setting of Example~\ref{ex:intro} where $n=3$, $k=2$ and $z=1$. Let $A$ denote the data of the Master, drawn from $GF(5)$, and $R$ denote the random matrix with elements drawn uniformly at random from $GF(5)$ and independently from $A$. Let $M_\text{SS}=\begin{bmatrix}A&R\end{bmatrix}^T$, the encoding is given as follows
\begin{equation*}
C=VM_\text{SS}=\begin{bmatrix}
1 & 1\\
1 & 2\\
1 & 3
\end{bmatrix} 
\begin{bmatrix}
A\\R
\end{bmatrix}
\end{equation*}
The resulting $3$ shares $(A+R, A+2R,A+3R)$ form the secret sharing code. The addition and multiplication are element wise, e.g., for share $2$ each element of $R$ is multiplied by $2$ and added to the correspondent element in $A$.
The Master can decode the secret by contacting any $k=2$ workers, downloading their shares and decoding $A$ and $R$. Secrecy is ensured, because $A$ is padded with $R$ in each share.
\end{example}

\subsection{Staircase codes}
\begin{table} \label{tab:unim}
\normalsize
\linespread{1}
\centering
\begin{tikzpicture}[baseline=(current  bounding  box.center)]
\tikzstyle{stealth} = [draw=none,text=black]

\def\x{-0.035}
\def\xshift{0.17}
\draw[xshift=\xshift cm,draw=none,fill=xgray]  (-1.04+\x,-1.38) -- (-1.04+\x,-0.96) -- (-0.182+\x,-0.96) -- (-.182+\x,-.49) -- (.677+\x,-.49) -- 
(.677+\x,-.3) -- (1.491+\x,-0.3) -- (1.491+\x,0) -- (2.75,0) -- (2.75,-1.38) -- cycle; 

\node[stealth] at (0,0)
{$M_\text{SC}=\left[\!
\begin{array}{c:c:c:c:c} 
 ~ & ~ &\multirow{2}{*}{$\cd_2$}  & \dots & \cd_{h-1}\\
 ~& \multirow{2}{*}{$\cd_1$} & & & \multirow{2}{*}{$\ck_h$} \\
\multirow{2}{*}{$\mathcal{S}$} & & \multirow{2}{*}{$\ck_3$}  &  \dots  & \\ 
 & \multirow{2}{*}{$\ck_2$} & & & \multirow{3}{*}{$\mathbf{0}$} \\
\multirow{2}{*}{$\ck_1$} &  & \multirow{2}{*}{$\mathbf{0}$} & \dots & \\
 & \mathbf{0} & & & \\
 \end{array}\!\right].$};
 
 \node[stealth] at (4.2,-0.5) {\Large $\substack{\text{staircase}\\ \text{structure}}$};
 \draw[->] (3.5,-0.6) ..controls (3,-1).. (2.4,-0.9);

\def\y{0.1}
\draw[xshift=\xshift cm,thin] (-1.9+\x,-1.38) -- (-1.04+\x,-1.38) -- (-1.04+\x,-0.96) -- (-0.182+\x,-0.96) -- (-.182+\x,-.49) -- (.677+\x,-.49) -- 
(.677+\x,-.3) -- (1.491+\x,-0.3) -- (1.491+\x,0) -- (2.74,0); 
\draw[xshift=\xshift cm,fill=white,draw=white] (-1.1,-1.386) -- (2.6,-1.386) -- (2.6,-1.49) -- (-1.1,-1.49) -- cycle;
\draw[xshift=\xshift cm,thin, gray,decorate,decoration={brace,amplitude=5pt,mirror},yshift=-0.8pt]  (-1.93+\x,-1.5+\y) -- (-1.04+\x,-1.5+\y);
\node[xshift=\xshift cm,stealth,color=gray] at (-1.45+\x,-2+\y) {$M_1$};
\draw[xshift=\xshift cm,thin, gray,decorate,decoration={brace,amplitude=5pt,mirror},yshift=-0.8pt]   (-1.04+\x,-1.5+\y) -- (-0.182+\x,-1.5+\y) ;
\node[xshift=\xshift cm,stealth,color=gray] at (-0.6,-2+\y) {$M_2$};
\draw[xshift=\xshift cm,thin, gray,decorate,decoration={brace,amplitude=5pt,mirror},yshift=-0.8pt]  (-.182+\x,-1.5+\y) -- (.677+\x,-1.5+\y);
\node[xshift=\xshift cm,stealth,color=gray] at (0.3+\x,-2+\y) {$M_3$};
\node[xshift=\xshift cm,stealth,gray] at (1.1+\x,-2+\y) {$\dots$};
\draw[xshift=\xshift cm,thin, gray,decorate,decoration={brace,amplitude=5pt,mirror},xshift=0.4pt,yshift=-0.8pt]  (1.446,-1.5+\y) -- (2.7,-1.5+\y);
\node[xshift=\xshift cm,stealth,color=gray] at (2.12,-2+\y) {$M_h$};
\draw[xshift=\xshift cm,thin] (-1.11,-1.38) -- (2.7,-1.38);

 \end{tikzpicture}
 \caption{ The structure of the matrix $M_\text{SC}$ that contains the secret  and keys in the universal Staircase code construction \cite{BRIT17}. }
 \label{tab:unim}
\end{table}
Let $A$ and $R$ be as defined previously. An $(n,k,z)$ Universal\footnote{For the sake of brevity, we only describe Universal Staircase codes and shall refer to them as Staircase codes. We refer the interested reader to~\cite{BRIT17} for more information about other families of Staircase codes.} Staircase code~\cite{BR16,BRIT17} allows the Master to encode $A$ into $n$ shares and distribute them to $n$ workers. In addition to the properties of classical secret sharing, Staircase codes enjoy the {\em secret reconstruction with minimum communication cost} property. The Master can reconstruct the secret by contacting any set of $d$, $k\leq d\leq n$, workers and downloading a part of their shares. The information theoretic lower bound on the amount of information downloaded from each worker is referred to as communication cost $CC(d)$ and is given~\cite{HLKBtrans} by
\begin{equation}
CC(d)=\dfrac{k-z}{d-z}.
\end{equation}

\noindent{\em Encoding: }Let $V$ be an  $n\times n$ Vandermonde matrix defined over $GF(q)$. Let $M_\text{SC}$ be the matrix defined in Table~\ref{tab:unim} and detailed next. The encoding of Staircase codes consists of multiplying $V$ by $M_\text{SC}$ to obtain the matrix $C=VM_\text{SC}$. The $n$ rows of $C$ form the $n$ different shares.

To construct the matrix $M$ defined in Table~\ref{tab:unim}, an $(n,k,z)$ Staircase code requires dividing the data matrix $A$ into $b(k-z)$ matrices $A_1,\dots,A_{(k-z)b}$ each of dimension\footnote{If the number of rows in $A$ is not divisible by $b$, one can use zero padding or the representation of $A$ in a smaller field $GF(q_1)$ such that $q=q_1^b$.} $m/b(k-z) \times \ell$, where $b= \text{LCM}\{k+1,\dots,n-z\}$. Let $d_1=n, d_2=n-1, \dots, d_h = k$ denote the number of workers contacted by the Master, with $h=n-k+1$. Let $b_i \triangleq d_i -z$ for $i=1,\dots, h$. The data matrices are arranged in a $b_1m / (k-z) b\times \ell (k-z)b /b_1$ matrix $\mathcal{S}$. To ensure secrecy, the construction uses $zb$ matrices $R_1,\dots, R_{zb}$ of dimension $m/b(k-z) \times \ell$ each and with elements drawn independently and uniformly at random from $GF(q)$. The random matrices $R_1,\dots,R_{zb}$ are partitioned into $h$ matrices $\mathcal{R}_i$, $i=1,\dots,h,$ each of dimension $zm/(k-z)b \times \ell (k-z)b/b_ib_{i-1}$ with $b_0=1$. 

\noindent The matrix $M_\text{SC}$ is the concatenation of $h$ matrices $M_i,$ $i=1,\dots,h,$ shown\footnote{In~\eqref{eq:mi} the dimensions of the rows are scaled by $m/(k-z)b$ for clarity of presentation.} in \eqref{eq:mi}. Each matrix $M_i$ consists of the $b_i$ sub-tasks downloaded by the Master when decoding from $d_i$ workers, i.e., when there are $n-d_i$ stragglers.
\be
\label{eq:mi}
\normalsize
\linespread{1}
  \resizebox{0.93\textwidth}{!}{
\begin{tikzpicture}[baseline=(current  bounding  box.center)]
\tikzstyle{stealth} = [draw=none,text=black]
\node[stealth] (1) at (0,0){$
M_1=\hspace{0.3cm} \bbm
\multirow{2}{*}{$\cf$} \\
\\
 \ck_1\\
\ebm  \quad$,};

\normalsize
\linespread{1}

\node[stealth] (2) [right=0.3cm of 1] {$M_2=\hspace{0.3cm}\bbm
\cd_1 \\ \ck_2 \\ \mathbf{0} \ebm \quad,\ \dots \ ,$} ;
\node[stealth] (3) [right=0.3cm of 2] {$M_j=\hspace{0.3cm}\bbm
\cd_{j-1} \\ \ck_j \\ \mathbf{0} \ebm \quad, \  \dots \ ,$};
\node[stealth] (4) [right=0.3cm of 3] {$M_{h}=\hspace{0.3cm}\bbm
\cd_{h-1} \\ \ck_{h} \\ \mathbf{0} \ebm \quad.$};

\def\y{0.03}
\def\h{0.05}
\def\x{0.2}
\def\j{1.3}
\draw[<->,gray] (-0.08+0.15,-0.78-\y) to (0.82,-0.78-\y);
\node[stealth,gray] (4) at (0.45,-1-\y) {\scriptsize $\ell (k-z)b/b_1$};

\draw[<->,gray,xshift=0.03cm] (3.2+\h+1*\x,-0.78-\y) to (3.93+\h+1*\x,-0.78-\y);
\node[stealth,gray] at (3.6+\h+1*\x,-1) {\scriptsize $ \ell (k-z)b/b_1b_2$};

\draw[<->,gray,xshift=0.11cm] (7.75+\h,-0.78-\y) to (3.65+3.88+\h+\j,-0.78-\y);
\node[stealth,gray,xshift=0.11cm] at (7+\h+\j,-1) {\scriptsize $ \ell (k-z)b/b_{j-1}b_{j}$};

\draw[<->,gray,xshift=0.25cm] (11.13+\h+\j,-0.78-\y) to (11.12+1.1+\h+\j,-0.78-\y);
\node[stealth,gray,xshift=0.25cm] at (11.72+\h+\j,-1) {\scriptsize $\ell b/b_{h-1}$};

\draw[<->,gray] (-0.05,-0.7) to (-0.05,0.7);
\node[stealth,gray] at (-0.25,0) {\scriptsize $n$};
\draw[<->,gray] (0.85+\h,-.7) to (0.85+\h,0);
\draw[<->,gray] (0.85+\h,0) to (0.85+\h,0.7);
\node[stealth,gray] at (1+\h,-0.35) {\scriptsize $z$};
\node[stealth,gray] at (1.07+\h,0.35) {\scriptsize $b_1$};

\draw[<->,gray,xshift=0.03cm] (3.15+0.2,-0.7) to (3.15+0.2,0.7);
\node[stealth,gray,xshift=0.05cm] at (3.15,0) {\scriptsize $n$};
\draw[<->,gray,xshift=0.01cm] (3.3+0.75+\h+1*\x,-.7) to (3.3+0.75+\h+1*\x,-0.25);
\draw[<->,gray,xshift=0.01cm] (3.3+0.75+\h+1*\x,-0.25) to (3.3+0.75+\h+1*\x,0.25);
\draw[<->,gray,xshift=0.01cm] (4.05+\h+1*\x,0.25) to (4.05+\h+1*\x,0.7);
\node[stealth,gray,xshift=0.01cm] at (3.35+0.85+\h+1*\x,0) {\scriptsize $z$};
\node[stealth,gray,xshift=0.01cm] at (3.35+0.92+\h+1*\x,0.45) {\scriptsize $b_2$};
\node[stealth, gray,xshift=0.01cm] at (3.35+0.85+\h+1*\x,-0.45) {\scriptsize 1};

\draw[<->,gray,xshift=0.15cm] (6.15+0.2+\j,-0.7) to (6.15+0.2+\j,0.7);
\node[stealth,gray,xshift=0.15cm] at (6.15+\j,0) {\scriptsize $n$};
\draw[<->,gray,xshift=0.12cm] (7.45+\h+1*\x+\j,-.7) to (7.45+\h+1*\x+\j,-0.25);
\draw[<->,gray,xshift=0.12cm] (7.45+\h+1*\x+\j,-0.25) to (7.45+\h+1*\x+\j,0.25);
\draw[<->,gray,xshift=0.12cm] (7.45+\h+1*\x+\j,0.25) to (7.45+\h+1*\x+\j,0.7);
\node[stealth,gray,xshift=0.12cm] at (7.45+0.15+\h+1*\x+\j,0) {\scriptsize $z$};
\node[stealth,gray,xshift=0.12cm] at (7.45+0.21+\h+1*\x+\j,0.45) {\scriptsize $b_j$};
\node[stealth, gray,xshift=0.12cm] at (7.45+0.45+\h+1*\x+\j,-0.45) {\tiny $n-d_j$};

\draw[<->,gray,xshift=0.25cm] (10.85+0.2+1*\j,-0.7) to (10.85+0.2+1*\j,0.7);
\node[stealth,gray,xshift=0.25cm] at (10.85+1*\j,0) {\scriptsize $n$};
\draw[<->,gray,xshift=0.22cm] (12.4+\h+1*\j,-.7) to (12.4+\h+1*\j,-0.25);
\draw[<->,gray,xshift=0.22cm] (12.4+\h+1*\j,-0.25) to (12.4+\h+1*\j,0.25);
\draw[<->,gray,xshift=0.22cm] (12.4+\h+1*\j,0.25) to (12.4+\h+1*\j,0.7);
\node[stealth,gray,xshift=0.22cm] at (12.4+0.15+\h+1*\j,0) {\scriptsize $z$};
\node[stealth,gray,xshift=0.22cm] at (12.4+0.21+\h+1*\j,0.45) {\scriptsize $b_h$};
\node[stealth, gray,xshift=0.22cm] at (12.4+0.25+0.15+\h+1*\j,-0.45) {\tiny $h-1$};

\end{tikzpicture}
}
\ee

The elements appearing in each matrix $\cd_j$ are the elements of the $\left(n-j+1\right)^{th}$ row of $\bbm M_1 \ M_2 \dots M_j \ebm$ rearranged to obtain the dimension of $\cd_j$ as $mb_{j+1}/ (k-z) b \times \ell (k-z)b/b_j b_{j+1}$  for $j=1,\dots, h-1$. The $\mathbf{0}$'s are the all zero matrices used to complete the $M_{i}$'s to $nm(k-z)b$ rows. 

The structure of the matrix $M_\text{SC}$, called Staircase structure, allows the Master to decode the secret and achieve optimal communication and read overheads $\CO$ and $\RO$ for all $d$, $k\leq d \leq n$.

\noindent{\em Decoding:} The Master contacts any $d_i$ workers, $i=1,\dots,h$ and downloads $\bbm M_1 & \dots & M_i \ebm$ from each contacted worker. The Master is guaranteed \cite[Theorem~2]{BRIT17} to decode the secret. 

In the setting of secure distributed computing, the Master encodes $A$ and sends the $n$ shares to the workers. To compute $A\mathbf{x}$, $\m$ sends $\mathbf{x}$ to the workers and waits for the first $d$, $k\leq d\leq n$, workers to send part of their shares multiplied by $\mathbf{x}$. Since the multiplication is linear, the Master can decode $A\mathbf{x}$ and part of the random matrices $R_i\mathbf{x}$, $i=1,\dots,zb$.

The following Example illustrates the encoding and decoding of Staircase codes.

\begin{example}[Staircase codes]
\label{ex:SC}
Consider the same setting of Example~\ref{ex:intro2}, where $n=3$, $k=2$ and $z=1$. Let the $m \times \ell$ matrix $A$ denote the Mater's data. Staircase code uses two matrices $R_1$ and $R_2$ to ensure secrecy and requires dividing $A$ into $2$ matrices $A_1,A_2$, each of dimension $m/2 \times \ell$. The matrices $\mathcal{S}$, $\mathcal{D}$, $\mathcal{R}_1$ and $\mathcal{R}_2$ are defined as $\mathcal{S}=\bbm A_1 & A_2\ebm^T$, $\mathcal{R}_1=R_1$, $\mathcal{D}=R_1$, $\mathcal{R}_2=R_2$. The encoding is given as follows
\be
\label{eq:encex}
\begin{tikzpicture}[baseline=(current  bounding  box.center)]
\tikzstyle{stealth} = [draw=none,text=black]

\node[stealth] (1) at (0,0)
{$\bbm A_1 &  R_1\\
A_2 & R_2\\ 
R_1 & 0\\ \ebm$};

\node[stealth] (3) [right=-0.2cm of 1] {.};
\node[stealth] (2) [left=-0.1cm of 1]  {$\bbm 1&1&1\\
1&2&4\\
1&3&4
 \ebm$};




\node[stealth] [left=1.8cm of 1] {$C=VM_\text{SC}=$};
%
%
 \end{tikzpicture}
\ee

The resulting shares are given in Table~\ref{tab:CESS}. The Master can decode the secret by either contacting any $k=2$ workers, downloading their shares and decoding $A$ and $R$, or contacting any $3$ workers, downloading half of their shares and decoding $A$ and $R_1$. Secrecy is ensured, because the secret is padded with the key in each share, c.f., Example~\ref{ex:intro2}.
\end{example}

\section{Bounds on the Master's mean waiting time for all $(n,k,z)$ systems} 
\label{sec:proof1}
We derive an upper and a lower bound on the Master's mean waiting time $\E[T_\text{SC}(n,k,z)]$ for all $(n,k,z)$ systems, i.e., we prove Theorem~\ref{thm:ublb}. We restate Theorem~\ref{thm:ublb} for the sake of presentation. 

\setcounter{temp}{\value{theorem}}
\setcounter{theorem}{0}
\begin{theorem}[Bounds on the Master's mean waiting time $\E {[T_\text{SC}] }$]
Let $H_n$ be the $n^{\text{th}}$ harmonic sum defined as $H_n \triangleq \sum_{i=1}^n \frac{1}{i}$, 
with the notation $H_0 \triangleq 0$. 
The mean waiting time of the Master $\mathbb{E}[T_{\text{SC}}]$ for an $(n,k,z)$ Staircase coded system is upper bounded by
\begin{align*}
\E[T_{\text{SC}}] &\leq \min_{d\in\{k,\dots,n\}}\left\{\frac{H_n - H_{n-d}}{\lambda(d-z)}+\dfrac{c}{d-z}\right\},
\end{align*}
and lower bounded by 
\begin{align*}
\E[T_{\text{SC}}] &\geq \frac{c}{n-z}+  \max_{d\in \{k,\dots,n\}}
\sum_{i=0}^{k-1}\binom{n}{i}\sum_{j=0}^{i}\binom{i}{j}\frac{2(-1)^j}{\lambda\left(2(n-i+j)(d-z)+(n-d)(n-d+1)\right)}.
\end{align*}
\end{theorem}
\setcounter{theorem}{\value{temp}}
We divide the proof into two parts: proof of the upper bound, and proof of the lower bound.
\subsection{Proof of the upper bound on the mean waiting time} 
\begin{IEEEproof}
We use Jensen's inequality to upper bound the mean waiting time $\E[T_{\text{SC}}]$. Since $\min$ is a convex function, we can use Jensen's inequality to upper bound the mean waiting time, 
\begin{align}
\label{eq:jensen}
\E[T_{\text{SC}}] &= \E\left[\min_{d\in\{k,\dots,n\}} \left\{\alpha_dT'_{(d)}+\frac{c}{d-z}\right\}\right] 
\leq \min_{d\in\{k,\dots,n\}}\left\{\alpha_d\E\left[ T'_{(d)}\right]+\frac{c}{d-z}\right\}.
\end{align}
We need the following Theorem in order to derive an exact expression of the mean of the $d^\text{th}$ order statistic of $n$ \emph{iid} exponential random variables.
\begin{theorem*}[Renyi~\cite{Renyi53}]
\label{thm:Renyi}
The $d^{\text{th}}$ order statistic $T'_{(d)}$ of $n$ \emph{iid} exponential random variables $T'_i$ 
is equal to the following random variable in the distribution 
\begin{align*}
T'_{(d)} &\triangleq \sum_{j=1}^{d}\frac{T'_j}{n-j+1}. 
\end{align*}
\end{theorem*}
Using Renyi's Theorem, the mean of the $d^{\text{th}}$ order statistic $\E \left[ T'_{(d)} \right]$ can be written as
\begin{align}
\label{eq:avg}
\E[T'_{(d)}]&=\E[T'_j]\sum_{j=0}^{d-1}\dfrac{1}{n-j}
=\dfrac{H_n-H_{n-d}}{\lambda(k-z)}.
\end{align}
From equations \eqref{eq:jensen} and \eqref{eq:avg}, the mean waiting time is upper bounded by
\begin{equation*}
\E[T_{\text{SC}}]\leq \min_{d\in\{k,\dots,n\}}\left\{\dfrac{H_n-H_{n-d}}{\lambda(d-z)}+\frac{c}{d-z}\right\}.
\end{equation*}
We give an intuitive behavior of the upper bound. The harmonic number can be approximated by $H_n\approx \log(n)+\gamma,$ where $\gamma \approx 0.577218$ is called the Euler-Mascheroni constant. 
Alternatively, we can use the upper and lower bounds $\log(n)< H_n < \log(n+1)$ on the Harmonic number $H_n$, 
to upper bound the mean waiting time 
\begin{align}\label{eq:finavg}
\mathbb{E}[T_{\text{SC}}]<\min 
&\left\{\min_{d\in\{k,\dots,n-1\}}\left\{\dfrac{1}{\lambda(d-z)}\log\left(\dfrac{n+1}{n-d}\right)+\frac{c}{d-z}\right\},\dfrac{1}{\lambda(n-z)}\log\left(n+1\right)+\frac{c}{n-z}\right\}.
\end{align}
\end{IEEEproof}
\subsection{Proof of the lower bound on the mean waiting time}
\begin{IEEEproof}
Recall that $T_{SC}=  \min\{\alpha_d T_{(d)}: d \in \{k,\dots,n\} \}=\min\{\alpha_d T'_{(d)}+\dfrac{c}{d-z}: d \in \{k,\dots,n\} \}$. Since the minimum of the sum is greater than the sum of the minimums, 
we can lower bound the waiting time $T_{\text{SC}}$ in terms of residual waiting time $T'_{\text{SC}} \triangleq  \min\{\alpha_dT'_{(d)}: d \in \{k,\dots,n\} \}$, as 
\eq{
T_{\text{SC}} =\min_{d\in \{k,\dots,n\}}\{\alpha_dT'_{(d)}+\dfrac{c}{d-z} \} &\geq T'_{\text{SC}} + \frac{c}{(n-z)}.
}
Since the mean of a continuous random variable can be computed by integrating the tail probability,  
we lower bound $\E[T'_{\text{SC}}]$ by lower bounding the tail probability of $T'_{SC}$ exceeding any threshold value $t$. 
We observe that $T'_{\text{SC}}$ is greater than $t$, 
if and only if the $d^{\text{th}}$ order statistic $T'_{(d)}$'s is greater than $\frac{t}{\alpha_d}$ for each $d\in\{k,\dots,n\}$. 
That is, 
\eq{
\{T'_{\text{SC}} > t\} &= \bigcap_{d=k}^n\left\{T'_{(d)} > \frac{t}{\alpha_d} \right\}. 
}
Recall that $t\alpha_d^{-1}(k-z) = t(d-z)$ is increasing in $d$, and so is $T'_{(d)}$. 
For the residual service times $T'_1,\dots, T'_n$, we consider the following set  
\eq{
\cC_d(t) &\triangleq \left\{ T'_{(k)} > \frac{t}{\alpha_d}\right\}\bigcap_{i=d+1}^n\left\{T'_{(i)}-T'_{(i-1)} > \frac{t}{\alpha_i} - \frac{t}{\alpha_{i-1}}\right\}.
}
For each $d \in \{k, \dots, n\}$, we observe that $\cC_d(t) \subseteq \{ T'_{\text{SC}} > t\}$  since $\{T'_{(k)} > t\alpha_d^{-1}\} \subseteq \cap_{j=k}^d\{T'_{(j)} > t \alpha_j^{-1}\}$. 
It follows that, $\Pr \left\{T'_{\text{SC}}>t\right\} \geq 
\max_{d \in \{k,\dots,n\}}\Pr(\cC_d(t))$. 
Next, we evaluate $\Pr(\cC_d(t))$ explicitly. 
To this end, we first observe that ${\alpha_j}^{-1}-{\alpha_{j-1}}^{-1}={(k-z)}^{-1}$ identically for each $j \in \{1, \dots, n\}$. 
Further, we apply Renyi's Theorem and independence of residual times $T'_i$s
to write 
\begin{align}
\label{eq:ub1}
\Pr\left(\cC_d(t) \right) &= \Pr\left\{T'_{(k)} > \frac{t}{\alpha_d} \right\}\prod_{j=d+1}^{n}\Pr\left\{\frac{T'_j}{n-j+1} > \frac{t}{(k-z)}\right\}.
\end{align}
In the following, we would use $F(t)= 1-e^{-\lambda t}$ for $t \geq 0$ to represent the cumulative distribution function (CDF) and $\bar{F}(t) = 1 - F(t)$ to represent the complementary cumulative distribution function (CCDF), of an exponential random variable with rate $\lambda$. 
It follows that the CCDF for the residual service time $T'_i$ is $\Pr\{T'_j > t \}= \bar{F}((k-z)t)$. 
Utilizing the exponential form, we can write 
\begin{align}
\label{eq:ub4}
\prod_{j=d+1}^{n}\Pr\left\{\frac{T'_j}{n-j+1} > \frac{t}{(k-z)}\right\} 
&=\bar{F}\left(\sum_{j=d+1}^{n}{(n-j+1)}t\right)=\bar{F}\left(\dfrac{(n-d)(n-d+1)t}{2}\right).
\end{align}
From definition, it follows that $\alpha_k = 1$. 
Further, the $k^{\text{th}}$ order statistic of $n$ residual service times exceeds a threshold if and only if at most $k-1$ different residual service times are less than the threshold, c.f., Lemma~\ref{lem:OrderStatDist}.  
That is, 
\begin{align}
\label{eq:ub5}
\Pr\left\{T'_{(k)}> t\right\} 
&=\sum_{i=0}^{k-1}\binom{n}{i}F\left({(k-z)t} \right)^i \bar{F}\left({(k-z)t}\right)^{n-i}.
\end{align}
Since $F(t) = 1 - \bar{F}(t)$, using the binomial expansion, we have
\begin{equation}
\label{eq:ub6}
F\left({(k-z)t}\right)^i=
\sum_{j=0}^{i}\binom{i}{j}(-1)^{j}\bar{F}\left({(k-z)t}\right)^{j}.
\end{equation}
Exploiting the exponential form of $\bar{F}(t)$, aggregating results from~\eqref{eq:ub4},~\eqref{eq:ub5} and~\eqref{eq:ub6}, we can re-write~\eqref{eq:ub1} as 
\begin{align}
\label{eq:ub8}
\Pr\left(\cC_d(t)\right) &=
\sum_{i=0}^{k-1}\binom{n}{i}\sum_{j=0}^{i}\binom{i}{j}(-1)^{j}\bar{F}\big( t(n-i+j)(d-z) +t (n-d)(n-d+1)/2\big).
\end{align}
The proof follows from the integral $\int_{0}^{\infty}e^{-xt}dt=\frac{1}{x}$, the linearity of integrals, 
and the following lower bound 
\eq{
\E[T'_{\text{SC}}]=\int_{0}^{\infty}\Pr \left\{T'_{\text{SC}}>t\right\}dt\geq \int_{0}^{\infty}\max_{d \in \{k, \dots, n\}} \Pr(\cC_d(t))dt \geq \max_{d \in \{k, \dots, n\}}\int_{0}^{\infty}\Pr(\cC_d(t))dt.
}
\end{IEEEproof}
\begin{lemma} 
\label{lem:OrderStatDist}
Marginal complementary distribution of $d^\text{th}$ order statistics $T'_{(d)}$ of $n$ \emph{iid} random variables $(T'_1, \dots, T'_n)$ with common distribution $f_{T'}(t)$ is given by 
\eq{
\Pr\{T'_{(d)} > t\} &= \sum_{i=0}^{d-1}\binom{n}{i}F_{T'}(t)\bar{F}_{T'}(t)^{n-i}.
}
\end{lemma}
We note the cumulative distribution function (CDF) of $f$ by $F_{T'}(t)\triangleq f_{T'}(T'<t)$ and the complementary cumulative distribution function (CCDF) of $f$ by $\bar{F}\triangleq f_{T'}(T'>t)= 1-F_{T'}(t)$.
\begin{IEEEproof}
The $d^\text{th}$ order statistic is greater than $t$, if and only if at most $d-1$ out of $n$ \emph{iid} random variables $(T_1,\dots, T_n)$ can be less than $t$, and the rest are greater than $t$. 
\end{IEEEproof}

\section{Distribution of the Master's waiting time for all $(n,k,z)$ systems}   
\label{sec:proof2}
Now we are ready to derive an integral expression for the probability distribution of $T_{\text{SC}}$, the Master's waiting time when using Staircase codes. 

\begin{theorem}[Integral expression leading to $F_{T_{\text{SC}}}(t)$]
\label{thm:exactexp}
The distribution of the Master's waiting time $T_{\text{SC}}$ of an $(n,k,z)$ system using Staircase codes is given by
\begin{equation}\label{eq:exactexp}
F_{T_{\text{SC}}}\left( t \right)=
1-n!\int_{(y_k, \dots, y_n)\in A(t)}\frac{F_{T'}(y_k)^{k-1}}{(k-1)!}dF_{T'}(y_k)\dots dF_{T'}(y_{n}) \quad \text{ for }t>0.
\end{equation}
We denote the residual service time at each worker $\w_i,\ i=1,\dots,n,$ by the random variable $T'_i = T_i - \frac{c}{k-z}$, 
and the associated distribution by $F(y_i)\triangleq F_{T'}(y_i)=1- \exp(-\lambda y_i)$ for $y_i > 0$.  
For $i=k,\dots,n$, we define $t_i $ as \mbox{$t_i\triangleq \max\left\{\Big(\frac{i-z}{k-z}\Big)\big(t -\frac{c}{i-z}\big), 0\right\}$}. We denote by $A(t)$ the set of ordered variables $(y_k, \dots, y_n)$ such that
\eq{
A(t) &\triangleq \{ 0 \leq y_k \leq y_{k+1} \leq \dots \leq y_n: t_k < y_k, \dots, t_n <  y_n \}.
} 

\end{theorem}

We apply Theorem~\ref{thm:exactexp} to get the mean waiting time of the Master and the exact distribution of the waiting time for systems with $n=k+1$ and $n=k+2$ in Theorem~\ref{Corr:MeanWaiting} and Corollary~\ref{Corr:ExactWaiting}, respectively.
\begin{corollary}[Exact expression of $F_{T_\text{SC}}(t)$ for systems with up to $2$ stragglers]
\label{Corr:ExactWaiting} 
The distribution of the Master's waiting time for $(k+1,k,z)$ and $(k+2,k,z)$ systems is given in \eqref{eq:corr3} and~\eqref{eq:corr4}, respectively.
\begin{align}
\label{eq:corr3}
F_{T_{\text{SC}}(k+1,k,z)}(t) &=  F_{T'}(t_{k+1})^{k+1}+F_{T'}(t_k)^k\bar{F}_{T'}(t_{k+1})(k+1). \\
F_{T_{\text{SC}}(k+2,k,z)}(t) &= F_{T'}(t_{k+2})^{k+2} + (k+2)\bar{F}_{T'}(t_{k+2})\Big[F_{T'}(t_{k+1})^{k+1} + (k+1)F_{T'}(t_k)^k(\bar{F}_{T'}(t_{k+1}) -\frac{1}{2}\bar{F}_{T'}(t_{k+2}))\Big].\label{eq:corr4}
\end{align}
Both distributions are defined for $t>0$, and $F_{T'}(t)\triangleq1-\exp(-\lambda(k-z)t)$.
\end{corollary}

We omit the proof of Corollary~\ref{Corr:ExactWaiting} since it follows from simply integrating~\eqref{eq:exactexp} and defer the proof of Theorem~\ref{Corr:MeanWaiting} to the Appendix.

\begin{IEEEproof}[Proof of Theorem~\ref{thm:exactexp}] Let $T'_i$ denote the residual service time of worker $i$ with the offset $\frac{c}{k-z}$. 
The sequence $(T'_1, \dots, T'_n)$ of residual service times of $n$ workers is assumed to be \emph{iid} and distributed exponentially with rate $\lambda(k-z)$ with the tail-distribution function $\bar{F}_{T'}(t) \triangleq e^{-\lambda (k-z) t}$ for $t > 0$.  
 
Since the common distribution of residual service times is absolutely continuous with respect to the Lebesgue measure, 
the corresponding probability density exists and is denoted by $f_{T'}(t) =dF_{T'}(t)/dt = \lambda(k-z)e^{-\lambda(k-z)t}$ for $t \ge 0$. 
Further, we know that the order statistics $(T'_{(1)}, \dots, T'_{(n)})$ of residual times $(T'_1, \dots, T'_n)$ is identical for all their $n!$ permutations.  
Hence, for any $0 \leq y_1 \leq \ldots \leq y_n$, we can write $f_{T'_{(1)}, \dots, T'_{(n)}}(y_1,\ldots, y_n) = n! f_{T'_1,\dots,T'_n}(y_1,\dots,y_n)=n!\prod_{i=1}^nf_{T'}(y_i)$. The product form of joint density follows from the independence of the residual service times. 

In terms of $\alpha_j = \frac{k-z}{j-z}$, the order statistics of residual times $T'_{(j)}$, and the offset $\frac{c}{k-z}$, we can write
\eq{
\left\{T_{\text{SC}} > t \right\}&= \bigcap_{j = k}^n\left\{T'_{(j)} > \frac{t}{\alpha_j} - \frac{c}{j-z} \right\}.
} 
For each $k \leq j \leq n$, we define $t_j \triangleq \max\left\{\frac{t}{\alpha_j} - \frac{c}{j-z},0\right\},  y_{n+1} \triangleq \infty$, and $\hat{A}(t) \triangleq \cap_{j=k}^{n+1}\{t_j <  y_j \leq y_{j+1}\}\cap_{j=1}^{k-1}\{0 \leq y_j \leq y_{j+1} \}$.  
In terms of $t_j, y_{n+1}$ and $\hat{A}(t)$, 
we can write the tail distribution 
\eq{
\Pr\{T_{\text{SC}} > t\} = \int_{y \in \hat{A}(t)}dF_{T'_{(1)},\dots,T'_{(n)}}(y) = n!\int_{t_n}^{\infty}\cdots\int_{t_k}^{y_{k+1}}\prod_{i=k}^ndF_{T'}(y_i)\left(\int_{0}^{y_k}\cdots\int_{0}^{y_{2}}\prod_{i=1}^{k-1}dF_{T'}(y_i)\right).
} 

First, we compute the integral with respect to ordered non-negative real variables $(y_1, \dots, y_{k-1})$ over the region $B_{k-1} \triangleq \cap_{j=1}^{k-1}\{0 \leq y_j \leq y_{j+1} \}$, a projection of $\hat{A}(t)$ on $(k-1)$ dimensional space spanned by $(y_1, \dots, y_{k-1})$. 
\begin{claim}
\label{claim:int}
For each $k > 1$, we have 
$I_k \triangleq \int_{B_{k-1}} dF_{T'}(y_{k-1})\dots dF_{T'}(y_1) = \int_{0}^{y_k}\cdots\int_{0}^{y_{2}}\prod_{i=1}^{k-1}dF_{T'}(y_i) = \cfrac{F_{T'}(y_k)^{k-1}}{(k-1)!}$. 
\end{claim}
Since the projection of $\hat{A}(t)$ on $(n-k+1)$ dimensional space spanned by $(y_k, \dots, y_n)$ is equal to $A(t)$, it follows that the integration of the first part is equal to $n!\int_{(y_k, \dots, y_n) \in A(t)}dF_{T'}(y_n)\dots dF_{T'}(y_k)$, giving us the result.
\end{IEEEproof}

\begin{IEEEproof}[Proof of Claim~\ref{claim:int}]
We prove the claim by induction on the number of integration variables $k$. 
The base case of $k=2$ holds trivially true. 
We assume that the induction hypothesis holds true for some $k \geq 2$, and show that it holds true for $k+1$. 
This can be shown by writing the integral $I_{k+1}$ in $(k+1)$ integration variables $y_1, \dots, y_{k+1}$ in terms of the integral $I_k$, 
and evaluating the integral by substituting the induction hypothesis for $I_k$ as follows
\eq{
I_{k+1}
&=\int_{0}^{y_{k+1}}I_kdF_{T'}(y_k) =\int_{0}^{y_{k+1}} \frac{F_{T'}(y_k)^{k-1}}{(k-1)!}dF_{T'}(y_k).
}
\end{IEEEproof}


\section{Interplay between code design and latency} 
\label{sec:imp}
{
Universal Staircase codes allows the master to decode $A\mathbf{x}$ from any random number $d$ of workers, $k\leq d \leq n$. The downside is that the universal construction requires a large number of sub-tasks $b=\text{LCM}\{k-z+1,\dots,n-z\}$. In many applications, there may be an overhead associated with excessive divisions into sub-tasks. We show that we can reduce the number of sub-tasks at the expense of a small increase of the Master's waiting time. Using the so-called $\Delta$-Universal Staircase codes \cite{BRIT17} reduces the number of sub-tasks at the expense of limiting the Master to a set $\Delta\subseteq\{k,\dots,n\}$ of number of workers allowing the Master to decode $A\mathbf{x}$. In other words, the Master can decode $A\mathbf{x}$ by downloading enough information from any $d$ workers, $d\in \Delta$. It remains to prove that $d$ is concentrated around its mean. Hence, restricting $d$ to an interval $\Delta$ centered around its mean, leads to a reduction in the Master's waiting time. Figure~\ref{fig:concentration} depicts the concentration of $d$ around its average for a $(100,50,1)$ system simulated on MATLAB. Figure~\ref{fig:unidelta}, depicts the normalized difference between the mean waiting time of Universal Staircase codes and $\Delta$-Universal Staircase codes for $(n,n/2,1)$ systems with $\lambda = c= 1$ and $\Delta=\{d^*-1, d^*, d^*+1\}$, where $d^*$ is the value of $d$ that minimizes our upper bound in~\eqref{eq:main1}.}

\begin{figure}[h]
  \setlength\figureheight{0.35\textwidth}
  \setlength\figurewidth{0.4\textwidth}
\centering
\hfill
\begin{minipage}[t]{0.4\textwidth}
\centering
\resizebox{0.8\textwidth}{!}{
\definecolor{mycolor1}{rgb}{0.00000,0.44700,0.74100}%
\begin{tikzpicture}

\begin{axis}[%
width=0.951\figurewidth,
height=\figureheight,
at={(0\figurewidth,0\figureheight)},
scale only axis,
xmin=50,
xmax=100,
xlabel={Number of contacted workers $d$},
xlabel style = {font =\LARGE},
ymin=0,
ymax=700,
ylabel={Occurence},
ylabel style = {at = {(-0.05,0.5)}, font =\LARGE},
axis background/.style={fill=white},
title style={font=\bfseries},
legend style={legend cell align=left,align=left,draw=white!15!black}
]
\addplot[fill=mycolor1,fill opacity=0.6,draw=black,ybar interval,area legend] plot table[row sep=crcr] {%
x	y\\
50	19\\
51	25\\
52	35\\
53	42\\
54	64\\
55	71\\
56	103\\
57	115\\
58	149\\
59	179\\
60	221\\
61	236\\
62	340\\
63	348\\
64	344\\
65	360\\
66	420\\
67	447\\
68	485\\
69	529\\
70	471\\
71	509\\
72	511\\
73	448\\
74	458\\
75	489\\
76	388\\
77	387\\
78	349\\
79	305\\
80	249\\
81	247\\
82	176\\
83	140\\
84	98\\
85	94\\
86	57\\
87	31\\
88	29\\
89	17\\
90	6\\
91	5\\
92	2\\
93	1\\
94	0\\
95	1\\
96	0\\
97	0\\
98	0\\
99	0\\
100	0\\
};

\end{axis}
\end{tikzpicture}
}
\captionsetup{width=1.1\textwidth}
\caption{Histogram of the number of contacted workers for an $(n,k,z)=(100,50,1)$ system simulated on MATLAB over $10000$ iterations with $\lambda=1$ and $c=1$.}
\label{fig:concentration}
\end{minipage} \hfill %
\begin{minipage}[t]{0.4\textwidth}
\centering
\resizebox{0.8\textwidth}{!}{
\definecolor{mycolor1}{rgb}{0.53, 0.66, 0.42}
\definecolor{mycolor2}{rgb}{1.0, 0.65, 0.0}
\definecolor{mycolor3}{rgb}{0.8, 0.31, 0.36}
\begin{tikzpicture}
\begin{axis}[%
width=0.951\figurewidth,
height=\figureheight,
at={(0\figurewidth,0\figureheight)},
scale only axis,
xmin=8,
xmax=100,
ymin=0,
ymax=5,
ytick={0,1,2,3,4,5},
yticklabels={{  0\%},{ 1\%},{ 2\%},{ 3\%},{ 4\%},{ 5\%}},
xlabel={Number of workers $n$},
xlabel style = {font =\LARGE},
ylabel={Normalized difference},
ylabel style = {at = {(-0.01,0.5)}, font =\LARGE},
axis background/.style={fill=white},
title style={font=\bfseries},
legend style={legend cell align=left,align=left,draw=white!15!black}
]
\addplot [color=mycolor3,only marks,mark=asterisk,mark options={solid},mark size = 5]
  table[row sep=crcr]{%
4	0.544358032560435\\
8	1.98290497556907\\
10	2.56004090899042\\
20	3.54563679909723\\
30	3.57659773937007\\
40	3.34109186784089\\
50	3.2008182855303\\
60	2.98569154693503\\
70	2.8895595812504\\
80	2.72681119239527\\
90	2.63044784103703\\
100	2.49308653608194\\
};
\end{axis}
\end{tikzpicture}}
\captionsetup{width=1.1\textwidth}
\caption{Normalized difference between the mean waiting time of Universal Staircase codes and $\Delta$-universal Staircase codes for systems with rate $k/n=1/2$, $z=1$, $\lambda=1$ and $c=1$.}
\label{fig:unidelta}
\end{minipage}
\hfill
\end{figure}

Next, we prove that the number of workers $d$ that minimize the waiting time is concentrated around its average.
\begin{lemma}\label{lem:main}
For an $(n,k,z)$ system, the probability distribution of the distance between $d$ and its average is 
\begin{align*}
\Pr\{|d-\E [d]| > t\} \leq 2e^{-2t^2/{n(n-k)^2}}.
\end{align*}
\end{lemma}
We prove Lemma~\ref{lem:main} by showing that the number of workers $d$ that first finish the aggregate computation is concentrated around its mean, using McDiarmid's inequality. 
Recall that $d : \R_+^n \to \{k, \ldots, n\}$ is a function of the compute times $T_1, \ldots, T_n$. 
\eq{
d(T_1, T_2, \dots, T_n)\triangleq \arg\min\left\{\frac{k-z}{i-z}T_{(i)}:  i\in\{k,\dots,n\} \right\}.
}
\begin{claim} 
The number of workers $d$ that minimize the waiting time is a bounded difference function of compute times with constants $(n-k, \dots, n-k)$. 
That is, for each $i \in [n]$ taking $t,t^{i} \in \R_+^n$ such that $t_j = t^i_j$ for each $j \in [n] \setminus \{i\}$ and $t_i \neq t^i_i$,
\begin{equation}\label{eq:bd}
\sup\{|g(t) - g(t^i)|: t, t^i \in \R_+^n\} \leq {n-k}.
\end{equation}
\end{claim}
The claim follows from the fact that $d\in\{k,\dots,n\}$. 
We prove the tightness of~\eqref{eq:bd} using the following example.
\begin{example}
Consider the following realizations (ordered for simplicity) of $T_1,\dots,T_n$ of an $(n,k,z)$ system, such that $T_k = t_k$, $T_i < t_k$ for $i = 1, \dots, k-1$, and $T_i > (\frac{i-z}{k-z})t_k$ for $i = k+1, \dots, n$. 
The corresponding $g(T_1,\dots,T_n)$ is equal to $k$, because $T_k< (\frac{k-z}{i-z})T_i$ for all $k< i\leq n$. 
Next, consider the ordered variables $(T_1, \dots, T'_n)$ where $T_n$ changes to $T'_n \in (T_{n-1},  (\frac{n-z}{k-z})t_k)$ while the other $T_j$'s, $j\in\{1,\dots,n-1\}$, remain unchanged, then $g(T_1, \ldots, T'_i, \ldots, T_n)=n$. 
We observe that the set $(T_{n-1},  (\frac{n-z}{k-z})t_k)$ is not always empty since the condition $(\frac{k-z}{n-z-1})T_{n-1} > t_k$ only implies that $(\frac{n-z-1}{k-z})t_k < T_{n-1} < T_n < (\frac{n-z}{k-z})t_k$. 
Hence, there always exist a case where
\begin{equation*}
\sup_{\substack{t_1, \ldots, t_n \in \R^n\\t'_i \in \R}}|g(t_1, \ldots, t_i, \ldots, t_n) - g(t_1, \ldots, t'_i, \ldots, t_n)| = {n-k}.
\end{equation*}
\end{example}
%
\noindent Therefore, we can apply the McDiarmid's inequality to obtain the concentration bound on $d$.

\section{Simulations} 
\label{sec:simu}
{
We use the normalized difference between the mean waiting time of Staircase codes and classical secret sharing codes as a performance metric for Staircase codes. We refer to this metric as the savings. Using the result of Theorem~\ref{thm:ublb}, we can get a lower and an upper bound on the savings brought by Staircase codes. The lower bound on the savings is given in \eqref{eq:division}.
\begin{equation}\label{eq:division}
\dfrac{\E[T_{\text{SS}}]-\E[T_{\text{SC}}]}{\E[T_{\text{SS}}]} \geq 1- \min_{d\in\{k,\dots,n\}} \left\{\dfrac{(k-z)(\lambda c + H_n - H_{n-d})}{(d-z)(\lambda c + H_n - H_{n-k})}\right\}.
\end{equation}

To get an idea of the actual savings and the tightness of the bound in \eqref{eq:division}, we ran numerical simulations of the mean waiting time induced by the use of Staircase codes. 
By looking at \eqref{eq:division}, we notice that the bound depends on $\lambda$ and $c$ only through\footnote{Note that for $c=0$ we go to the exponential model and the savings would depend only on $\lambda$.} $\lambda c $ (our simulations show that the actual savings also have a strong dependency on $\lambda c$). Therefore, we consider three cases for $\lambda c$ :  large values of $\lambda c$ ($\lambda c=100$), medium values of $\lambda c$ ($\lambda c =1$) and small values of $\lambda c$ ($\lambda c =0.001$). We ran the simulations for two regimes:
\begin{itemize}
\item {\em Fixed rate $k/n$}: the plots can be seen in Figure~\ref{fig:rate}. We deduce from the plots that the lower bound is tighter for large values of $\lambda c$. Moreover,  the savings increase with the decrease of the rate $k/n$ and the increase of $\lambda c$. Note that for large values of $\lambda c$,  the lower bound in \eqref{eq:division} converges to $1-k/n$.
 
\item {\em Fixed number of parities $n-k$}: the plots can be seen in Figure~\ref{fig:pars}. We deduce from the plots that similarly to the fixed rate regime the lower bound is tight for large values of $\lambda c$ and that the savings increase with the increase of the number of parities $n-k$ and with the increase of $\lambda c$. However, we observe that the savings vanish asymptotically  with $n$ in this regime.
\end{itemize}
}

\vspace{-0.3cm} 

\begin{figure}[h!]
\centering
  \setlength\figureheight{0.35\textwidth}
  \setlength\figurewidth{0.4\textwidth}
\begin{minipage}[t]{0.33\textwidth}
\centering
\resizebox{0.8\textwidth}{!}{
\definecolor{mycolor1}{rgb}{0.0, 0.58, 0.71}
\definecolor{mycolor3}{rgb}{1.0, 0.65, 0.0}
\definecolor{mycolor4}{rgb}{0.8, 0.31, 0.36}
\definecolor{mycolor5}{rgb}{0.09, 0.45, 0.27}
\begin{tikzpicture}

\begin{axis}[%
width=0.951\figurewidth,
height=\figureheight,
at={(0\figurewidth,0\figureheight)},
scale only axis,
xmin=0,
xmax=100,
xlabel={Number of workers $n$},
xlabel style = {font=\large},
ymin=0,
ymax=0.9,
ytick={0,0.1,0.2,0.3,0.4,0.5,0.6,0.7,0.8,0.9,1,1.1,1.2,1.3,1.4},
yticklabels={{  0\%},{ 10\%},{ 20\%},{ 30\%},{ 40\%},{ 50\%},{ 60\%},{ 70\%},{ 80\%},{ 90\%},{100\%},{110\%},{120\%},{130\%},{140\%}},
ylabel={Percentage of time savings},
ylabel style = {font=\Large},
axis background/.style={fill=white},
title style={font=\bfseries,align=center},
legend style={at = {(1,0.4)},legend cell align=left,align=left,draw=white!15!black}
]
\addplot [color=mycolor3,ultra thick,solid,mark=o,mark options={solid},mark size = 3]
  table[row sep=crcr]{%
4	0.661708065796966\\
6	0.592720542115149\\
8	0.562554449610315\\
10	0.545501124242523\\
14	0.526577896181399\\
20	0.51257180270713\\
24	0.507044144710536\\
40	0.495725101861329\\
50	0.492224745498086\\
60	0.489810254107768\\
80	0.486734073778828\\
100	0.484817292512321\\
};
\addlegendentry{Simulation for $k/n=1/2$};

\addplot [color=mycolor1,ultra thick,solid,mark=square,mark options={solid},mark size = 3]
  table[row sep=crcr]{%
8	0.853649690686601\\
12	0.813047255174988\\
16	0.793801223432746\\
20	0.78251246697517\\
24	0.775050293525529\\
40	0.760195337930246\\
48	0.756456889673416\\
52	0.75500984967369\\
60	0.752726713233282\\
80	0.748942011623746\\
100	0.746659858446033\\
};
\addlegendentry{Simulation for $k/n=1/4$};

\addplot [color=mycolor4,ultra thick,dashed,mark=o,mark options={solid},mark size = 3]
  table[row sep=crcr]{%
4	0.661695664181165\\
6	0.592711611727679\\
8	0.562556296695627\\
10	0.54547250708937\\
14	0.52657281711827\\
20	0.512533898667835\\
24	0.506996866274248\\
40	0.495411583786386\\
50	0.491640509312192\\
60	0.488971770305604\\
80	0.485351296820639\\
100	0.482933911197609\\
};
\addlegendentry{Bound in \eqref{eq:division} for $k/n=1/2$};

\addplot [color=mycolor5,ultra thick,dashed,mark=square,mark options={solid},mark size = 3]
  table[row sep=crcr]{%
8	0.853652207098334\\
12	0.81305230683451\\
16	0.793810754932203\\
20	0.78250741043969\\
24	0.775031914011398\\
40	0.760037635832298\\
48	0.756214893915218\\
52	0.754725938821079\\
60	0.752312803207177\\
80	0.748270341173316\\
100	0.745727571724628\\
};
\addlegendentry{Bound in \eqref{eq:division} for $k/n=1/4$};

\end{axis}
\end{tikzpicture}
}
\captionsetup{subtype,width=0.95\textwidth}
\caption{Savings for $\lambda c=100$.}
\label{fig:savings1}
\end{minipage}%
\begin{minipage}[t]{0.33\textwidth}
\centering
\resizebox{0.8\textwidth}{!}{
\definecolor{mycolor1}{rgb}{0.0, 0.58, 0.71}
\definecolor{mycolor3}{rgb}{1.0, 0.65, 0.0}
\definecolor{mycolor4}{rgb}{0.8, 0.31, 0.36}
\definecolor{mycolor5}{rgb}{0.09, 0.45, 0.27}
\begin{tikzpicture}

\begin{axis}[%
width=0.951\figurewidth,
height=\figureheight,
at={(0\figurewidth,0\figureheight)},
scale only axis,
xmin=0,
xmax=100,
xlabel={Number of workers $n$},
xlabel style = {font=\large},
ymin=0,
ymax=0.9,
ytick={0,0.1,0.2,0.3,0.4,0.5,0.6,0.7,0.8,0.9,1,1.1,1.2,1.3,1.4},
yticklabels={{  0\%},{ 10\%},{ 20\%},{ 30\%},{ 40\%},{ 50\%},{ 60\%},{ 70\%},{ 80\%},{ 90\%},{100\%},{110\%},{120\%},{130\%},{140\%}},
ylabel={Percentage of time savings},
ylabel style = {font=\Large},
axis background/.style={fill=white},
title style={font=\bfseries,align=center},
legend style={legend cell align=left,align=left,draw=white!15!black}
]

\addplot [color=mycolor3,ultra thick,solid,mark=o,mark options={solid},mark size = 3]
  table[row sep=crcr]{%
4	0.431598987663955\\
6	0.318658049745731\\
8	0.268139412015294\\
10	0.239163291093237\\
14	0.203755702363028\\
20	0.175745892286212\\
24	0.164353990838816\\
40	0.13884743592206\\
50	0.130094274596354\\
60	0.123914009310282\\
80	0.114910994409288\\
100	0.109261097504249\\
};
\addlegendentry{Simulation for $k/n=1/2$};

\addplot [color=mycolor1,ultra thick,solid,mark=square,mark options={solid},mark size = 3]
  table[row sep=crcr]{%
8	0.685416296620852\\
12	0.594747793093936\\
16	0.549908184093677\\
20	0.522849201601285\\
24	0.504377631748719\\
40	0.46595730385185\\
48	0.456049740989562\\
52	0.451948131586827\\
60	0.445414232388386\\
80	0.434408733046191\\
100	0.427335522088201\\
};
\addlegendentry{Simulation for $k/n=1/4$};

\addplot [color=mycolor4,ultra thick,dashed,mark=o,mark options={solid},mark size = 3]
  table[row sep=crcr]{%
4	0.350877192982456\\
6	0.242268041237114\\
8	0.185870356882739\\
10	0.156567570360674\\
14	0.128545560907276\\
20	0.10898719687699\\
24	0.102744459697922\\
40	0.0892513332543982\\
50	0.0853911711564682\\
60	0.0828327892661047\\
80	0.0797329089284137\\
100	0.0779578308806478\\
};
\addlegendentry{Bound in \eqref{eq:division} for $k/n=1/2$};

\addplot [color=mycolor5,ultra thick,dashed,mark=square,mark options={solid},mark size = 3]
  table[row sep=crcr]{%
8	0.650140845070422\\
12	0.55466140082668\\
16	0.509547185873547\\
20	0.483518981071277\\
24	0.467370089879397\\
40	0.435076584993183\\
48	0.4270960974912\\
52	0.424172582683075\\
60	0.419407377007188\\
80	0.411708335206285\\
100	0.407164333059922\\
};
\addlegendentry{Bound in \eqref{eq:division} for $k/n=1/4$};

\end{axis}
\end{tikzpicture}}
\captionsetup{subtype,width=0.95\textwidth}
\caption{Savings for $\lambda c=1$.}
\label{fig:savings2}
\end{minipage}%
\begin{minipage}[t]{0.33\textwidth}
\centering
\resizebox{0.8\textwidth}{!}{
\definecolor{mycolor1}{rgb}{0.0, 0.58, 0.71}
\definecolor{mycolor3}{rgb}{1.0, 0.65, 0.0}
\definecolor{mycolor2}{rgb}{0.54, 0.2, 0.14}
\definecolor{mycolor4}{rgb}{0.8, 0.31, 0.36}
\definecolor{mycolor5}{rgb}{0.09, 0.45, 0.27}
\begin{tikzpicture}

\begin{axis}[%
width=0.951\figurewidth,
height=\figureheight,
at={(0\figurewidth,0\figureheight)},
scale only axis,
xmin=0,
xmax=100,
xlabel={Number of workers $n$},
xlabel style = {font=\large},
ymin=0,
ymax=0.9,
ytick={0,0.1,0.2,0.3,0.4,0.5,0.6,0.7,0.8,0.9,1,1.1,1.2,1.3,1.4},
yticklabels={{  0\%},{ 10\%},{ 20\%},{ 30\%},{ 40\%},{ 50\%},{ 60\%},{ 70\%},{ 80\%},{ 90\%},{100\%},{110\%},{120\%},{130\%},{140\%}},
ylabel={Percentage of time savings},
ylabel style = {font=\Large},
axis background/.style={fill=white},
title style={font=\bfseries,align=center},
legend style={legend cell align=left,align=left,draw=white!15!black}
]
\addplot [color=mycolor3,ultra thick,solid,mark=o,mark options={solid},mark size = 3]
  table[row sep=crcr]{%
4	0.292366011461477\\
6	0.189331075026696\\
8	0.144077161850632\\
10	0.118763366631336\\
14	0.0897697239964039\\
20	0.0669155696989134\\
24	0.0575736002301377\\
40	0.0385696065589682\\
50	0.0321710475416646\\
60	0.0274129089988584\\
80	0.0217454723598463\\
100	0.0178479460815239\\
};
\addlegendentry{Simulation for $k/n=1/2$};

\addplot [color=mycolor1,ultra thick,solid,mark=square,mark options={solid},mark size = 3]
  table[row sep=crcr]{%
8	0.43161578778841\\
12	0.300660202388475\\
16	0.238258256125676\\
20	0.20033790280004\\
24	0.175363771385976\\
40	0.120650184156797\\
48	0.106226524343054\\
52	0.0990684555663046\\
60	0.0895064445640014\\
80	0.0730407921604821\\
100	0.0621994754499431\\
};
\addlegendentry{Simulation for $k/n=1/4$};

\addplot [color=mycolor4,ultra thick,dashed,mark=o,mark options={solid},mark size = 3]
  table[row sep=crcr]{%
4	0.0721620079863089\\
6	0\\
8	0\\
10	0\\
14	0\\
20	0\\
24	0\\
40	0\\
50	0\\
60	0\\
80	0\\
100	0\\
};
\addlegendentry{Bound in \eqref{eq:division} for $k/n=1/2$};

\addplot [color=mycolor5,ultra thick,dashed,mark=square,mark options={solid},mark size = 3]
  table[row sep=crcr]{%
8	0.21206754044161\\
12	0.0710851774377032\\
16	0.0255982402956515\\
20	0.0098705675248304\\
24	0.00240808961088224\\
40	1.33226762955019e-15\\
48	1.22124532708767e-15\\
52	1.11022302462516e-15\\
60	1.11022302462516e-15\\
80	1.22124532708767e-15\\
100	1.22124532708767e-15\\
};
\addlegendentry{Bound in \eqref{eq:division} for $k/n=1/4$};

\end{axis}
\end{tikzpicture}}
\captionsetup{subtype,width=0.95\textwidth}
\caption{Savings for $\lambda c=0.001$.}
\label{fig:savings3}
\end{minipage}%
\caption{Savings for the fixed rate regime, $k/n = 1/2$ and $1/4$. The lower bound on the savings of Staircase codes obtained from \eqref{eq:division} is compared to the numerical values obtained by simulations. We consider systems with no colluding workers, i.e.,  $z=1$, we fix $\lambda =1$ and vary $c$. For instance, for systems with rate $k/n=1/2$ and $\lambda c = 100$ Staircase codes can provide up to $66\%$ reduction in the mean waiting time.}
\label{fig:rate}
\end{figure}

\vspace{-0.5cm}

\begin{figure}[h!]
\centering
  \setlength\figureheight{0.35\textwidth}
  \setlength\figurewidth{0.4\textwidth}
\begin{minipage}[t]{0.33\textwidth}
\centering
\resizebox{0.8\textwidth}{!}{
\definecolor{mycolor1}{rgb}{0.0, 0.58, 0.71}
\definecolor{mycolor3}{rgb}{1.0, 0.65, 0.0}
\definecolor{mycolor2}{rgb}{0.54, 0.2, 0.14}
\definecolor{mycolor4}{rgb}{0.8, 0.31, 0.36}
\definecolor{mycolor5}{rgb}{0.09, 0.45, 0.27} 

\begin{tikzpicture}

\begin{axis}[%
width=0.951\figurewidth,
height=\figureheight,
at={(0\figurewidth,0\figureheight)},
scale only axis,
xmin=8,
xmax=100,
xlabel={Number of workers $n$},
xlabel style = {font=\large},
ymin=0,
ymax=0.91,
ytick={0,0.1,0.2,0.3,0.4,0.5,0.6,0.7,0.8,0.9,1,1.1,1.2,1.3,1.4},
yticklabels={{  0\%},{ 10\%},{ 20\%},{ 30\%},{ 40\%},{ 50\%},{ 60\%},{ 70\%},{ 80\%},{ 90\%},{100\%},{110\%},{120\%},{130\%},{140\%}},
ylabel={Percentage of time savings},
ylabel style = {font=\Large},
axis background/.style={fill=white},
title style={font=\bfseries,align=center},
legend style={legend cell align=left,align=left,draw=white!15!black}
]
\addplot [color=mycolor3,ultra thick,solid,mark=o,mark options={solid},mark size = 3]
  table[row sep=crcr]{%
8	0.707798191493572\\
10	0.545481110072909\\
14	0.370681579444461\\
20	0.246571333205876\\
24	0.199831451647668\\
40	0.109209397608871\\
50	0.0830569157138005\\
60	0.0658452060219856\\
80	0.0450138183641426\\
100	0.0329810733899744\\
};
\addlegendentry{Simulation for $n-k=5$};

\addplot [color=mycolor1,ultra thick,solid,mark=square,mark options={solid},mark size = 3]
  table[row sep=crcr]{%
12	0.906428761165495\\
16	0.656930609874228\\
20	0.512527831832339\\
24	0.418434935893039\\
40	0.235438552787093\\
48	0.190874001783283\\
52	0.173974515485421\\
60	0.146950440315231\\
80	0.103896358257236\\
100	0.0786251031986798\\
};
\addlegendentry{Simulation for $n-k=10$};

\addplot [color=mycolor4,ultra thick,dashed,mark=o,mark options={solid},mark size = 3]
  table[row sep=crcr]{%
8	0.707790129623828\\
10	0.54547250708937\\
14	0.370698846520788\\
20	0.246551607423924\\
24	0.199784542038955\\
40	0.108688542118939\\
50	0.0819818949884217\\
60	0.0643365714630468\\
80	0.0424616313868847\\
100	0.0294367486205254\\
};
\addlegendentry{Bound in \eqref{eq:division} for $n-k=5$};

\addplot [color=mycolor5,ultra thick,dashed,mark=square,mark options={solid},mark size = 3]
  table[row sep=crcr]{%
12	0.906432842158472\\
16	0.65694734722281\\
20	0.512533898667835\\
24	0.418366612571501\\
40	0.234920765629741\\
48	0.190055552597478\\
52	0.172904706613768\\
60	0.145584779585891\\
80	0.101510735431523\\
100	0.0752605035565546\\
};
\addlegendentry{Bound in \eqref{eq:division} for $n-k=10$};

\end{axis}
\end{tikzpicture}}
\captionsetup{subtype,width=0.95\textwidth}
\caption{Savings for $\lambda c=100$.}
\label{fig:savingsk1}
\end{minipage}%
\begin{minipage}[t]{0.33\textwidth}
\centering
\resizebox{0.8\textwidth}{!}{
\definecolor{mycolor1}{rgb}{0.0, 0.58, 0.71}
\definecolor{mycolor3}{rgb}{1.0, 0.65, 0.0}
\definecolor{mycolor2}{rgb}{0.54, 0.2, 0.14}
\definecolor{mycolor4}{rgb}{0.8, 0.31, 0.36}
\definecolor{mycolor5}{rgb}{0.09, 0.45, 0.27}
\begin{tikzpicture}

\begin{axis}[%
width=0.951\figurewidth,
height=\figureheight,
at={(0\figurewidth,0\figureheight)},
scale only axis,
xmin=8,
xmax=60,
xlabel={Number of workers $n$},
xlabel style = {font=\large},
ymin=0,
ymax=0.9,
ytick={0,0.1,0.2,0.3,0.4,0.5,0.6,0.7,0.8,0.9,1,1.1,1.2,1.3,1.4},
yticklabels={{  0\%},{ 10\%},{ 20\%},{ 30\%},{ 40\%},{ 50\%},{ 60\%},{ 70\%},{ 80\%},{ 90\%},{100\%},{110\%},{120\%},{130\%},{140\%}},
ylabel={Percentage of time savings},
ylabel style = {font=\Large},
axis background/.style={fill=white},
title style={font=\bfseries,align=center},
legend style={legend cell align=left,align=left,draw=white!15!black}
]
\addplot [color=mycolor3,ultra thick,solid,mark=o,mark options={solid},mark size = 3]
  table[row sep=crcr]{%
8	0.444519595063966\\
10	0.238610297543876\\
14	0.0973969331131002\\
20	0.0400631721049633\\
24	0.0260865160085009\\
40	0.00835068623363181\\
50	0.00522805484508618\\
60	0.00351934173285017\\
80	0.00196931084572093\\
100	0.00128807936715018\\
};
\addlegendentry{Simulation for $n-k=5$};

\addplot [color=mycolor1,ultra thick,solid,mark=square,mark options={solid},mark size = 3]
  table[row sep=crcr]{%
12	0.780215834087443\\
16	0.340055723006408\\
20	0.176119478056586\\
24	0.103350768151847\\
40	0.0249314592520378\\
48	0.0156147743268842\\
52	0.0128366918099953\\
60	0.00908794553552483\\
80	0.00467623222941605\\
100	0.00284144869688428\\
};
\addlegendentry{Simulation for $n-k=10$};

\addplot [color=mycolor4,ultra thick,dashed,mark=o,mark options={solid},mark size = 3]
  table[row sep=crcr]{%
8	0.381576763485477\\
10	0.156567570360674\\
14	0.0207873843550476\\
20	0\\
24	0\\
40	0\\
50	0\\
60	0\\
80	0\\
100	0\\
};
\addlegendentry{Bound in \eqref{eq:division} for $n-k=5$};

\addplot [color=mycolor5,ultra thick,dashed,mark=square,mark options={solid},mark size = 3]
  table[row sep=crcr]{%
12	0.758367895545315\\
16	0.280686736968939\\
20	0.10898719687699\\
24	0.0396430603053308\\
40	0\\
48	0\\
52	0\\
60	0\\
80	0\\
100	0\\
};
\addlegendentry{Bound in \eqref{eq:division} for $n-k=10$};

\end{axis}
\end{tikzpicture}}
\captionsetup{subtype,width=0.95\textwidth}
\caption{Savings for $\lambda c=1$.}
\label{fig:savingsk2}
\end{minipage}%
\begin{minipage}[t]{0.33\textwidth}
\centering
\resizebox{0.8\textwidth}{!}{
\definecolor{mycolor1}{rgb}{0.0, 0.58, 0.71}
\definecolor{mycolor3}{rgb}{1.0, 0.65, 0.0}
\definecolor{mycolor2}{rgb}{0.54, 0.2, 0.14}
\definecolor{mycolor4}{rgb}{0.8, 0.31, 0.36}
\definecolor{mycolor5}{rgb}{0.09, 0.45, 0.27}
\begin{tikzpicture}

\begin{axis}[%
width=0.951\figurewidth,
height=\figureheight,
at={(0\figurewidth,0\figureheight)},
scale only axis,
xmin=8,
xmax=60,
xlabel={Number of workers $n$},
xlabel style = {font=\large},
ymin=0,
ymax=0.5,
ytick={0,0.1,0.2,0.3,0.4,0.5,0.6,0.7,0.8,0.9,1,1.1,1.2,1.3,1.4},
yticklabels={{  0\%},{ 10\%},{ 20\%},{ 30\%},{ 40\%},{ 50\%},{ 60\%},{ 70\%},{ 80\%},{ 90\%},{100\%},{110\%},{120\%},{130\%},{140\%}},
ylabel={Percentage of time savings},
ylabel style = {font=\Large},
axis background/.style={fill=white},
title style={font=\bfseries,align=center},
legend style={legend cell align=left,align=left,draw=white!15!black}
]
\addplot [color=mycolor3,ultra thick,solid,mark=o,mark options={solid},mark size = 3]
  table[row sep=crcr]{%
8	0.241991516012818\\
10	0.118967972103229\\
14	0.0497817840108536\\
20	0.0218914500619507\\
24	0.0148357311958503\\
40	0.00533149558466131\\
50	0.00347797571963716\\
60	0.00239762045011274\\
80	0.00141275398319013\\
100	0.000936713020812326\\
};
\addlegendentry{Simulation for $n-k=5$};

\addplot [color=mycolor1,ultra thick,solid,mark=square,mark options={solid},mark size = 3]
  table[row sep=crcr]{%
12	0.483253086761915\\
16	0.135340023677621\\
20	0.0666678494553131\\
24	0.040836482970572\\
40	0.0120676796050147\\
48	0.00807604233904916\\
52	0.00679415886378366\\
60	0.00506115722786254\\
80	0.00280592406670081\\
100	0.00180373065340052\\
};
\addlegendentry{Simulation for $n-k=10$};

\addplot [color=mycolor4,ultra thick,dashed,mark=o,mark options={solid},mark size = 3]
  table[row sep=crcr]{%
8	0.0271885705955242\\
10	0\\
14	0\\
20	0\\
24	0\\
40	0\\
50	0\\
60	0\\
80	0\\
100	0\\
};
\addlegendentry{Bound in \eqref{eq:division} for $n-k=5$};

\addplot [color=mycolor5,ultra thick,dashed,mark=square,mark options={solid},mark size = 3]
  table[row sep=crcr]{%
12	0.270505504640035\\
16	0\\
20	0\\
24	0\\
40	0\\
48	0\\
52	0\\
60	0\\
80	0\\
100	0\\
};
\addlegendentry{Bound in \eqref{eq:division} for $n-k=10$};

\end{axis}
\end{tikzpicture}}
\captionsetup{subtype,width=0.95\textwidth}
\caption{Savings for $\lambda c=0.001$.}
\label{fig:savingsk3}
\end{minipage}%
\caption{Savings for the fixed number of parities regime, $n-k=5$ and $10$. The lower bound on  the savings of Staircase codes obtained from \eqref{eq:division} is compared to the numerical values obtained by simulations. Similarly to Figure~\ref{fig:rate}, we consider systems with $z=1$, $\lambda =1$ and vary $c$.}
\label{fig:pars}
\end{figure}

 \begin{table}[h!]
\newcolumntype{L}[1]{>{\raggedright\let\newline\\\arraybackslash\hspace{0pt}}m{#1}}
\newcolumntype{C}[1]{>{\centering\let\newline\\\arraybackslash\hspace{0pt}}m{#1}}
\newcolumntype{R}[1]{>{\raggedleft\let\newline\\\arraybackslash\hspace{0pt}}m{#1}}
\centering
\def\arraystretch{1.2}
\begin{tabular}{|c|c|c|c|c|c|c|}
\hline
\multicolumn{1}{|c}{\bf Size of matrix $A$ }& \multicolumn{1}{c}{$\mathbf{(n,k,z)}$} & \multicolumn{1}{c}{$\mathbf{c^*}$} &\multicolumn{1}{c}{$\mathbf{\lambda^*}$} & \multicolumn{1}{c}{\bf Lower bound in~\eqref{eq:division}} & \multicolumn{1}{c}{\bf Savings from simulations} & \multicolumn{1}{c|}{\textbf{Savings on EC2}}\\ \hline
\multirow{3}{*}{\begin{varwidth}{2cm} \centering $387000 \times 250$ (large)\end{varwidth}} & $(4,2,1)$ &  $0.8380$ & $0.7996$ & $ 30\%$ & $39\%$ & ${59\%}$ \\ \cline{2-7}
& $(10,5,1)$ & $0.4317$ & $1.6996$ & $ 12\%$ & $21\%$ &\bf $ {42\%}$\\ \cline{2-7}
& $(20,10,1)$ & $0.3478$ & $2.0938$ & $ 7\%$ &$14\%$ & \bf ${32\%}$\\ \hline
\multirow{3}{*}{\begin{varwidth}{2cm} \centering $42000 \times 250$ (small)\end{varwidth}} & $(4,2,1)$ &  $0.1090$ & $8.2783$ & $ 33\%$ & $42\%$ & \bf ${42\%}$\\ \cline{2-7}
 & $(4,2,1)$ &  $0.0267$ & $36.7524$ & $ 35\%$ & $43\%$ &\bf  ${30\%}$\\ \cline{2-7}
 & $(4,2,1)$ &  $0.1641$ & $5.8613$ & $ 34\%$ & $43\%$ &\bf  ${25\%}$\\ \hline
\end{tabular}
\caption{Comparison of the performance of Staircase codes on Amazon EC2 to the  theoretical bound in~\eqref{eq:division} and the value obtained by simulations assuming the shifted exponential model in Section~\ref{sec:sysmod}. The shift $c^*$ and the rate $\lambda^*$ of the workers are measured, respectively, as the minimum response time and the inverse of the average response time at one worker over $1000$ iterations. }
\label{tab:boundaws}
\end{table}

\section{Implementation and Validation of the Theoretical Model}\label{sec:AWS}
{We describe a representative sample of our implementation on Amazon EC2 clusters and discuss our observations. In Section~\ref{sec:awshalf}, we present traces for systems with fixed rate $k/n=1/2$ (Figure~\ref{fig:amzdiffn}). We noticed that the straggler behavior, and therefore the savings, can depend on the date and time of the implementation. To highlight this dependence, we present in Section~\ref{sec:aws4} the traces of one system implemented at different date and times (Figure~\ref{fig:amzdifftime}).

{\em Discussion on the theoretical model:} Before giving the details, we summarize our findings in Table~\ref{tab:boundaws} that lists all the parameters that we implemented and compares the savings obtained on Amazon to the theoretical lower bound~\eqref{eq:division} and numerical savings obtained by simulations. We observe that the savings of the system on EC2 can surpass the numerical values resulting from our theoretical model in Section~\ref{sec:sysmod} for large sizes of the matrix $A$. However,  for small sizes of $A$, the savings in practice can be less. 

The difference between the theoretical results and the implementations can be attributed to several reasons.  First, in our model we assume  in \eqref{eq:sub} that the total service time of a  task does not change when divided into $b$ sub-tasks, each  requiring the same service time. Whereas, our implementation on Amazon shows that the  download time decreases faster than linearly with the size of the sub-task for large sub-tasks. Second, for small sub-tasks, we noticed an additional  overhead of sending the results of multiple sub-tasks. This  overhead becomes non-negligible when the task is small. Third, we have assumed a homogeneous setting where all workers have the same behavior which is not always the case in practice.  

Despite these differences, our adopted theoretical model is more amenable to theoretical analysis and provides insightful engineering guiding principles.}

 \begin{figure}[h!]
  \setlength\figureheight{0.35\textwidth}
  \setlength\figurewidth{0.4\textwidth}
\centering
\begin{minipage}[t]{0.3\textwidth}
\centering
\resizebox{!}{0.9\textwidth}{
\definecolor{mycolor1}{rgb}{0.09, 0.45, 0.27} 
\definecolor{mycolor2}{rgb}{0.8, 0, 0}
\definecolor{mycolor3}{rgb}{0.8, 0.31, 0.36}

} 
\captionsetup{subtype,width=\textwidth}
\caption{$(n,k,z)=(4,2,1)$.}
\label{fig:amzwaiting42}
\end{minipage}  \hfill%
\begin{minipage}[t]{0.3\textwidth}
\centering
\resizebox{!}{0.9\textwidth}{
\definecolor{mycolor2}{rgb}{0.8, 0, 0}
\definecolor{mycolor1}{rgb}{0.09, 0.45, 0.27} 
%
}
\captionsetup{subtype,width=\textwidth}
\caption{$(n,k,z)=(10,5,1)$.}
\label{fig:amzwaiting105}
\end{minipage}\hfill%
\begin{minipage}[t]{0.3\textwidth}
\centering
\resizebox{!}{0.9\textwidth}{
\definecolor{mycolor2}{rgb}{0.8, 0, 0}
\definecolor{mycolor1}{rgb}{0.09, 0.45, 0.27} 
%
}
\captionsetup{subtype,width=\textwidth}
\caption{$(n,k,z)=(20,10,1)$.}
\label{fig:amzwaiting2010}
\end{minipage} \hfill%
\caption{Empirical complementary CDF of the Master's waiting time (and its average) observed on Amazon EC2 clusters for systems with rate $k/n=1/2$. The data matrix $A$ is a $378000 \times 250$ matrix with entries generated uniformly at random from $\{1,\dots,255\}$. Staircase codes bring $59\%$ reduction in the mean waiting time for $n=4$. Those numbers were obtained by repeating the multiplication process $1000$ times.}
\label{fig:amzdiffn}
\end{figure}

\newpage
 
\subsection{Implementation for systems with rate $k/n=1/2$}\label{sec:awshalf}
We present the implementation of $(4,2,1)$, $(10,5,1)$ and $(20,10,1)$ systems on Amazon EC2 clusters. We use M4.large EC2 instances \cite{EC2} from Amazon web services (AWS) for our implementation. We assign the Master's job to an instance located in Virginia and the workers job to instances located in Ohio. We plot in Figures~\ref{fig:amzwaiting42},~\ref{fig:amzwaiting105} and~\ref{fig:amzwaiting2010} the empirical complementary CDF of the Master's waiting time for Staircase codes and classical secret sharing codes for $(4,2,1)$, $(10,5,1)$ and $(20,10,1)$ systems, respectively. The average  savings brought by Staircase codes are $59\%$, $42\%$ and $32\%$ for systems with $n=4$, $n=10$ and $n=20$ workers, respectively. These results are also summarized in Table~\ref{tab:boundaws}.
Note that for this set of implementation, the Master's data $A$ is a matrix of size $378000\times 250$ with entries generated uniformly at random from $\{1,\dots,255\}$. We run $1000$ multiplications of $A$ by a randomly generated vector $\mathbf{x}$.

\begin{figure}[h!]
  \setlength\figureheight{0.35\textwidth}
  \setlength\figurewidth{0.4\textwidth}
\centering
\begin{minipage}[t]{0.3\textwidth}
\centering
\resizebox{!}{0.8\textwidth}{
\definecolor{mycolor1}{rgb}{0.09, 0.45, 0.27} 
\definecolor{mycolor2}{rgb}{0.8, 0, 0}
}
\captionsetup{subtype,width=0.95\textwidth}
\caption{Tuesday 10-10-2017, 12:15 PM.}
\label{fig:amzwaiting1}
\end{minipage} \hfill%
\begin{minipage}[t]{0.3\textwidth}
\centering
\resizebox{!}{0.8\textwidth}{
\definecolor{mycolor1}{rgb}{0.09, 0.45, 0.27} 
\definecolor{mycolor2}{rgb}{0.8, 0, 0}
%
}
\captionsetup{subtype,width=0.95\textwidth}
\caption{Saturday 01-20-2018, 12:10 PM.}
\label{fig:amzwaiting2}
\end{minipage} \hfill %
\begin{minipage}[t]{0.3\textwidth}
\centering
\resizebox{!}{0.8\textwidth}{
\definecolor{mycolor1}{rgb}{0.09, 0.45, 0.27} 
\definecolor{mycolor2}{rgb}{0.8, 0, 0}
%
}
\captionsetup{subtype,width=\textwidth}
\caption{Thursday 10-5-2017, 2:24 PM.}
\label{fig:amzwaiting3}
\end{minipage}  \hfill %
\caption{An $(n,k,z)=(4,2,1)$ system implemented on Amazon EC2 cluster at different times.  The matrix $A$ is a $42000 \times 250$ matrix representing $56$ images from the LFW dataset \cite{learned2016labeled}. We observe that the distribution of the Master's waiting time and  the savings brought by using Staircase codes ($42\%$, $30\%$, and $25\%$ respectively) depend on the date and time of the implementation.}
\label{fig:421}
\label{fig:amzdifftime}
\end{figure}

\subsection{Implementation on $4$ worker instances at different times}\label{sec:aws4}
We present the trace of a $(4,2,1)$ system implemented at different dates and times on Amazon EC2 clusters. We follow the same setting as before except that $A$ is a $42000 \times 250$ matrix generated using the LFW dataset of public faces\footnote{To obtain the data matrix $A$, we convert the first $56$ faces to $3$ matrices each. Each matrix is a $250\times 250$ matrix representing the color value of the pixels of each image in red, green and blue, respectively.} \cite{learned2016labeled}. We observe that the distribution of the Master's waiting time and the savings brought by using Staircase codes depend on the date and time of the implementation. This can be due to the varying state of the instances and the varying volume of traffic at Amazon servers. 

\section{Conclusion and open problems}
\label{sec:conc}

We consider the problem of secure coded computing. We propose the use of a new family of secret sharing codes called Staircase codes that reduces the delays caused by stragglers. We show that Staircase codes always lead to smaller waiting time compared to classical secret sharing codes, e.g., Shamir secret sharing codes. The reason behind reducing the delays is that Staircase codes allow flexibility in the number of stragglers up to a given maximum, and universally achieve the information theoretic limit on the download cost by the Master, leading to latency reduction. We consider the shifted exponential model for the workers's response time. In our analysis, we find upper and lower bounds on the Master's mean waiting time. We characterize the distribution of the Master's waiting time, and its mean, for systems with $n=k-1$ and $n=k-2$. For general $(n,k,z)$ systems. Moreover,  we derive an expression that can give the exact distribution, and the mean, of the waiting time of the Master. We supplement our theoretical study with extensive implementation on Amazon EC2 clusters.

While Staircase codes reduce the Master's waiting time by minimizing the download cost, they are not designed to minimize latency. The problem of designing codes that minimize the latency remains open in general. Another open problem, which we leave for future work, is when the malicious workers corrupt the results sent to the Master.

\appendices
\setcounter{temp}{\value{theorem}}
\setcounter{theorem}{1}
\section{Proof of Theorem~\ref{Corr:MeanWaiting}}\label{sec:proofcorr}
For the clarity of presentation, we restate Theorem~\ref{Corr:MeanWaiting}.
\begin{theorem}[Exact expression of $\E {[T_\text{SC}]} $ for systems with up to $2$ stragglers]
\label{Corr:MeanWaiting} 
The mean waiting time of the Master for $(k+1,k,z)$ and $(k+2,k,z)$ systems is given in \eqref{eq:corr1} and~\eqref{eq:corr2}, respectively.
\begin{align}
\label{eq:corr1}
\mathbb{E}\left[T_{\text{SC}}(k+1,k,z)\right] &= \frac{c}{k-z+1}+\frac{1}{\lambda}\sum_{i=1}^{k+1}(-1)^{i}\binom{k+1}{i}\left[\frac{i\exp\left(\frac{-\lambda c}{k-z}\right)}{(k-z)i+1}-\dfrac{1}{(k-z+1) i} \right].\\
\label{eq:corr2}
\E[T_{\text{SC}}(k+2,k,z)]&= \E[T_\text{SC}(k+2,k+1,z)]+\frac{1}{\lambda}\sum_{i=2}^{k+2}(-1)^i\binom{k+2}{i}\binom{i}{2}\left[\dfrac{\exp\left(-\frac{4 \lambda c}{k-z}\right)}{(k-z)i+4}-\dfrac{2\exp\left(-\frac{3 \lambda c}{k-z}\right)}{(k-z)i+3}\right].
\end{align}
\end{theorem}
\setcounter{theorem}{\value{temp}}
We derive the expression of the Master's mean waiting time for $(n,k,z)=(k+1,k,z)$ and $(n,k,z)=(k+2,k,z)$ systems. Applying Theorem~\ref{thm:exactexp} for the case of $n=k+1$, we get
\eq{
\bar{F}_{T_{\text{SC}}(k+1,k,z)}(t) &=  1- F_{T'}(t_{k+1})^{k+1}-F_{T'}(t_k)^k\bar{F}_{T'}(t_{k+1})(k+1), \quad \text{for } t>0.
}
Recall that $t_k$ and $t_{k+1}$ are defined as
$t_k = \max \left\{ t - \frac{c}{k-z}, 0 \right\}$ and $t_{k+1} = \max\left\{ \frac{k+1-z}{k-z}\left(t- \frac{c}{k+1-z}\right),0 \right\}$.

Since $F_{T'}(0) = 0$,  we can compute the Master's mean waiting time $\E \left[ T(k+1,k,z)\right]$ as 
\begin{align*}
\E \left[ T_{\text{SC}}(k+1,k,z)\right]&=\int_{0}^{\infty}(1- (1-\bar{F}_{T'}(t_{k+1}))^{k+1})dt - \int_{0}^{\infty}(1-\bar{F}_{T'}(t_k))^k\bar{F}_{T'}\left(t_{k+1}\right)(k+1) dt,\\
& = \frac{c}{k+1-z} + \int_{\frac{c}{k+1-z}}^{\infty}(1- (1-\bar{F}_{T'}(t_{k+1}))^{k+1})dt - \int_{\frac{c}{k-z}}^{\infty}(1-\bar{F}_{T'}(t_k))^k\bar{F}_{T'}\left(t_{k+1}\right)(k+1) dt.
\end{align*}
Using the binomial expansion and integrating the exponential function $\bar{F}_{T'}(t)=\exp(-\lambda(k-z)t)$, we get
\begin{align*}
\mathbb{E}\left[T_{\text{SC}}(k+1,k,z)\right] &= \frac{c}{k-z+1}+\frac{1}{\lambda}\sum_{i=1}^{k+1}(-1)^{i}\binom{k+1}{i}\left[\frac{i\exp\left(\frac{-\lambda c}{k-z}\right)}{(k-z)i+1}-\dfrac{1}{(k-z+1) i} \right].
\end{align*}
Similarly, we apply Theorem~\ref{thm:exactexp} for $n = k+2$ and get 
\eq{
\bar{F}_{T_{\text{SC}}(k+2,k,z)}(t)&=1 - F_{T'}(t_{k+2})^{k+2} - (k+2)\bar{F}_{T'}(t_{k+2})\Big[F_{T'}(t_{k+1})^{k+1} + (k+1)F_{T'}(t_k)^k(\bar{F}_{T'}(t_{k+1}) -\frac{1}{2}\bar{F}_{T'}(t_{k+2}))\Big].
}
Recall that for $i=k,k+1,k+2$, we define $t_i$ as $t_i\triangleq \max\left\{\dfrac{i-z}{k-z}\left(t-\frac{c}{i-z}\right)\right\}$.
Since $F_{T'}(0) = 0$,  we can compute the Master's mean waiting time $\E \left[T_{\text{SC}}(k+2,k,z)\right]$ as 
\begin{align*}
\E \left[T_{\text{SC}}(k+2,k,z)\right]&= \int_{0}^{\infty}(1 - F_{T'}(t_{k+2})^{k+2}) dt - \int_{0}^{\infty}(k+2)\bar{F}_{T'}(t_{k+2})F_{T'}(t_{k+1})^{k+1}dt \\
&~~ -\int_{0}^{\infty} (k+2)(k+1)\bar{F}_{T'}(t_{k+2})F_{T'}(t_k)^k\left(\bar{F}_{T'}(t_{k+1})-\frac{1}{2}\bar{F}_{T'}(t_{k+2}) \right)dt\\
&=\frac{c}{k+2-z}+\int_{\frac{c}{k+2-z}}^{\infty}(1- (1-\bar{F}_{T'}(t_{k+2}))^{k+2})dt - (k+2)\int_{\frac{c}{k+1-z}}^{\infty}(1-\bar{F}_{T'}(t_{k+1}))^{k+1}\bar{F}_{T'}\left(t_{k+2}\right) dt\\
&-\binom{k+2}{2}\int_{\frac{c}{k-z}}^{\infty}(1-\bar{F}_{T'}(t_k))^{k}\bar{F}_{T'}(t_{k+2})\left(2\bar{F}(_{T'}t_{k+1}) - \bar{F}_{T'}(t_{k+2})\right)dt.
\end{align*}
Using the binomial expansion and integrating the exponential function $\bar{F}_{T'}(t)=\exp(-\lambda(k-z)t)$, we get  
\eq{
\E\left[ T_{\text{SC}}(k+2,k,z)\right]&
=\frac{c}{k-z+2}+\sum_{i=1}^{k+2}\frac{(-1)^i\binom{k+2}{i}}{\lambda}\left[\dfrac{i\exp\left(-\frac{\lambda c}{k-z+1}\right)}{(k-z+1)i+ 1}-\dfrac{1}{(k-z+2)i}\right] \nonumber\\
&+\sum_{i=2}^{k+2}\frac{(-1)^i\binom{k+2}{i}\binom{i}{2}}{\lambda}\left[\dfrac{\exp\left(-\frac{4 \lambda c}{k-z}\right)}{i(k-z)+4}-\dfrac{2\exp\left(-\frac{3\lambda c}{k-z}\right)}{i(k-z)+3}\right].
}

\section{Hiding the attribute vectors}
\label{sec:hiding}

\begin{figure}[h!]
\centering
\begin{minipage}[t]{0.5\textwidth}
\centering
\resizebox{1.05\textwidth}{0.4\textwidth}{
 \begin{tikzpicture}[>=stealth', auto,
 triangle/.style = {fill=white, regular polygon, regular polygon sides=3 },]
  \definecolor{lightgray}{rgb}{0.83, 0.83, 0.83}
\def\mx{3}
\def\my{-4}
\tikzstyle{server} = [fill=black!10, rectangle, rounded corners=4mm, draw,minimum width=2em, minimum height=2.5em]

\node[inner sep=0] (s1) at (-4.3,0) {\includegraphics[height=9mm,width=7mm]{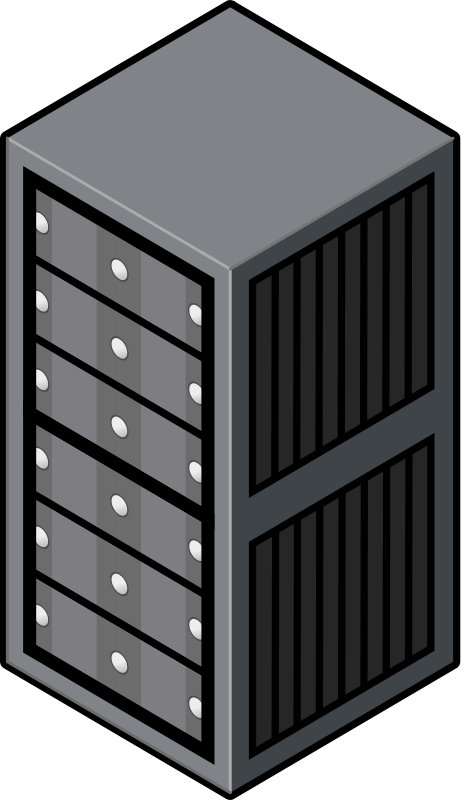}} node[below=0cm of s1,color=byzantium] {Master} node[above=0cm of s1,color=byzantium] {$\m$};


\node[server,right=0.8cm of s1,font=\footnotesize] (bb) {$\substack{\mbox{$(n_2,k_2,z_2)$}\\ \mbox{Staircase code}}$}; 
\node[above=0.8cm of bb, font=\footnotesize] (rm) {Randomness};

\node[inner sep=0pt,right =0.75 of bb,font=\footnotesize] (us2){\includegraphics[height=7mm]{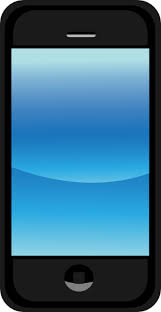}}
node[right =0 of us2,font=\small,color=byzantium] {$\w_{n_1+2}$};

\node[inner sep=0,above=0.35 cm of us2,font=\footnotesize] (us1) {\includegraphics[width=5mm]{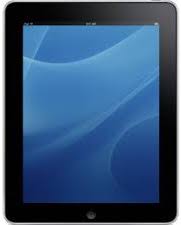}}
node[right= 0 cm of us1,font=\small,color=byzantium] {$\w_{n_1+1}$};

\node[below=-0.15cm of us2] (v1) {$\vdots$};
\node[above left=-0.35 and 0.3 of v1] {$\vdots$};

\node[inner sep=0,below= 0.7 cm of us2,font=\footnotesize] (us3) {\includegraphics[width=6mm]{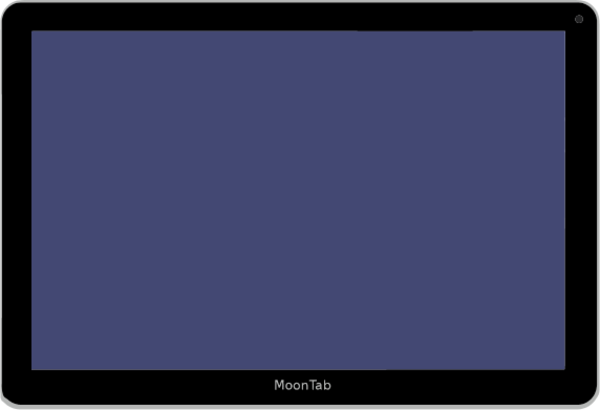}}
node[right=0 of us3,font=\small,color=byzantium] {$\w_{n}$} ;
 
\path[every node/.style={},color=black,->]
(s1) edge node[midway,below=0,font=\footnotesize] {$A$} node[midway,above=0,font=\footnotesize] {Data} (bb) 
   (bb.north east)    	edge  node[midway,above=0,sloped,font=\footnotesize] {$S_{n_1+1}$}  (us1) 
   (bb.east)		edge  node[midway,above=0,sloped,font=\footnotesize] {$S_{n_1+2}$}  (us2)
   (bb.south east)	edge  node[midway,below=0,sloped,font=\footnotesize] {$S_{n}$}  (us3)
   (rm) 			edge  node[midway,right=0,font=\footnotesize] {$R$} (bb);
   
\node[above=0 of us1,color=byzantium] {Group $2$};


\node[server,left=0.8cm of s1,font=\footnotesize] (bb) {$\substack{\mbox{$(n_1,k_1,z_1)$}\\ \mbox{Staircase code}}$}; 
\node[above=0.8cm of bb, font=\footnotesize] (rm) {Randomness};

\node[inner sep=0pt,left =0.75 of bb,font=\footnotesize] (us2){\includegraphics[height=7mm]{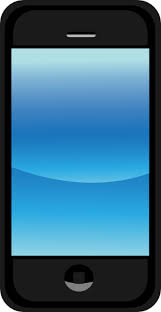}}
node[left =0 of us2,font=\small,color=byzantium] {$\w_{2}$};

\node[below=-0.15cm of us2] (v1) {$\vdots$};
\node[above right=-0.35 and 0.3 of v1] {$\vdots$};

\node[inner sep=0,above=0.35 cm of us2,font=\footnotesize] (us1) {\includegraphics[width=5mm]{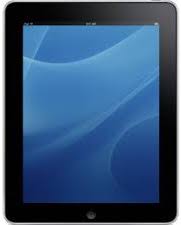}}
node[left= 0 cm of us1,font=\small,color=byzantium] {$\w_{1}$};

\node[inner sep=0,below= 0.7 cm of us2,font=\footnotesize] (us3) {\includegraphics[width=6mm]{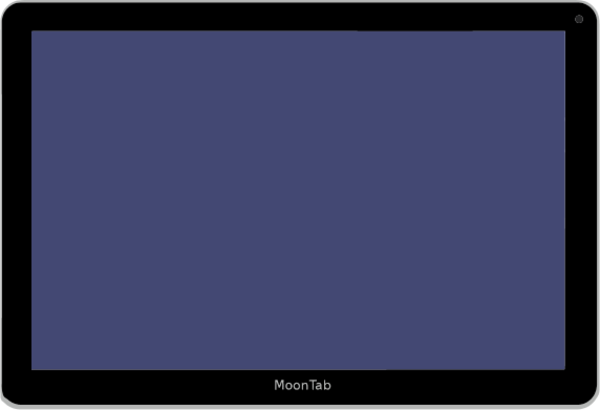}}
node[left=0 of us3,font=\small,color=byzantium] {$\w_{n_1}$} ;
 
\path[every node/.style={},color=black,->]
(s1) edge node[midway,below=0,font=\footnotesize] {$A$} node[midway,above=0,font=\footnotesize] {Data} (bb) 
   (bb.north west)    	edge  node[midway,above=0,sloped,font=\footnotesize] {$S_{1}$}  (us1) 
   (bb.west)		edge  node[midway,above=0,sloped,font=\footnotesize] {$S_{2}$}  (us2)
   (bb.south west)	edge  node[midway,below=0,sloped,font=\footnotesize] {$S_{n_1}$}  (us3)
   (rm) 			edge  node[midway,right=0,font=\footnotesize] {$R$} (bb);
   
\node[above=0 of us1,color=byzantium] {Group $1$};

\end{tikzpicture}}
\captionsetup{subtype,width=\textwidth}
\caption{$\m$ encodes $A$ using an $(n_1,k_1,z_1)$ and $(n_2,k_2,z_2)$ Staircase codes and distribute the obtained shares to the workers.}
\label{fig:apdxfig1}
\end{minipage} \hfill%
\begin{minipage}[t]{0.45\textwidth}
\centering
\resizebox{\textwidth}{!}{
\begin{tikzpicture}
\node[inner sep=0] (s) at (0.7,0) {\includegraphics[height=9mm,width=7mm]{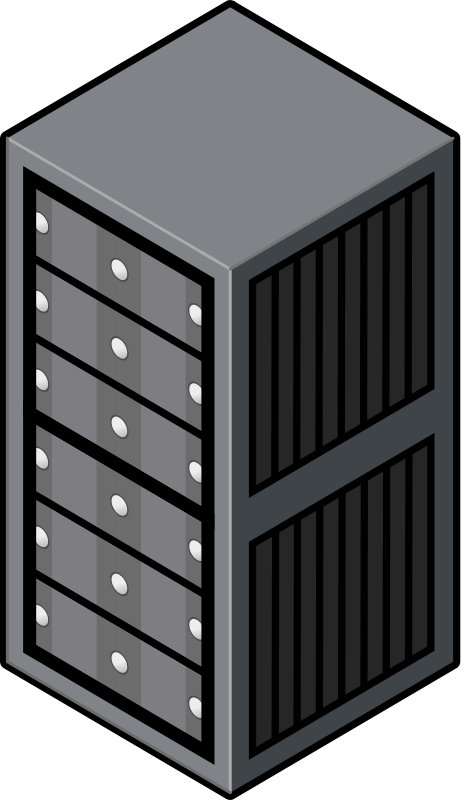}} node[below=0cm of s,color=byzantium] {$\m$};


\node[inner sep=0pt,right =3 of s,font=\footnotesize] (us2){\includegraphics[height=7mm]{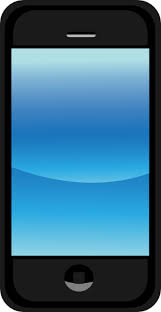}}
node[left =-0.1 of us2,font=\small,color=byzantium] {} node[below =0 of us2,font=\footnotesize] {$S_{n_1+2}$} node[right =0 of us2,font=\small,color=byzantium] {$\w_{n_1+2}$};;

\node[below=0.2 cm of us2] (v1) {$\vdots$};
\node[above left = -0.3 and 1.4 of v1] {$\vdots$};

\node[inner sep=0,above= 0.6 cm of us2,font=\footnotesize] (us1) {\includegraphics[width=5mm]{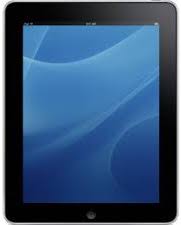}}
node[left=-0.1cm of us1,font=\small,color=byzantium] {} node[below=0 of us1,font=\footnotesize] {$S_{n_1+1}$} node[right =0 of us1,font=\small,color=byzantium] {$\w_{n_1+1}$};;

\node[inner sep=0,below=1 cm of us2,font=\footnotesize] (us3) {\includegraphics[width=6mm]{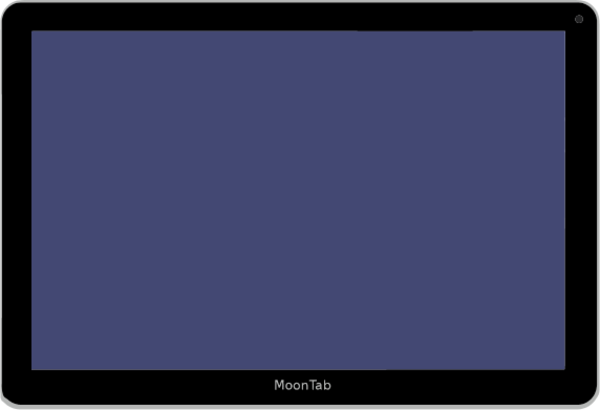}}
node[left=-0.1 of us3,font=\small,color=byzantium] {} node[below= 0cm of us3,font=\footnotesize] {$S_{n}$} node[right =0 of us3,font=\small,color=byzantium] {$\w_{n}$};;

 
\path[every node/.style={},color=black,->]
   (s.north east)    	edge[bend left]  node[midway,above left=0 and -0.4 cm,sloped,font=\footnotesize] {$\mathbf{u}$} (us1) 
   (s.east)		edge[bend left]  node[midway,below=0,sloped,font=\footnotesize] {$\mathbf{u}$}  (us2)
   (s.south east)		edge[bend right]  node[midway,below=0,sloped,font=\footnotesize] {$\mathbf{u}$}   (us3)
			
   (us1)		edge[dashed]  node[midway,above =0, sloped,font=\footnotesize]  {$S_{n_1+1}\mathbf{u}$}  	(s.north east) 
   (us2)		edge[dashed]  node[midway,below =0, sloped,font=\footnotesize]  {$S_{n_1+2}\mathbf{u}$}  	(s)
   (us3)		edge[dashed]  node[midway,below =0, sloped,font=\footnotesize]  {$S_{n}\mathbf{u}$}  	(s.south east)
   ;
   
\node[above=0 of us1,color=byzantium] {Group $2$};


\node[inner sep=0pt,left =3 of s,font=\footnotesize] (us2) {\includegraphics[height=7mm]{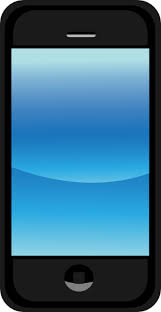}}
node[left =-0.1 of us2,font=\small,color=byzantium] {} node[below =0 of us2,font=\footnotesize] {$S_{2}$} node[left =0 of us2,font=\small,color=byzantium] {$\w_{2}$};;

\node[below=0.2 cm of us2] (v1) {$\vdots$};
\node[above right = -0.3 and 1.2 of v1] {$\vdots$};

\node[inner sep=0,above= 0.6 cm of us2,font=\footnotesize] (us1) {\includegraphics[width=5mm]{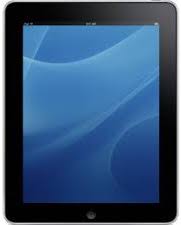}}
node[left=-0.1cm of us1,font=\small,color=byzantium] {} node[below=0 of us1,font=\footnotesize] {$S_{1}$} node[left =0 of us1,font=\small,color=byzantium] {$\w_{1}$};;

\node[inner sep=0,below=1 cm of us2,font=\footnotesize] (us3) {\includegraphics[width=6mm]{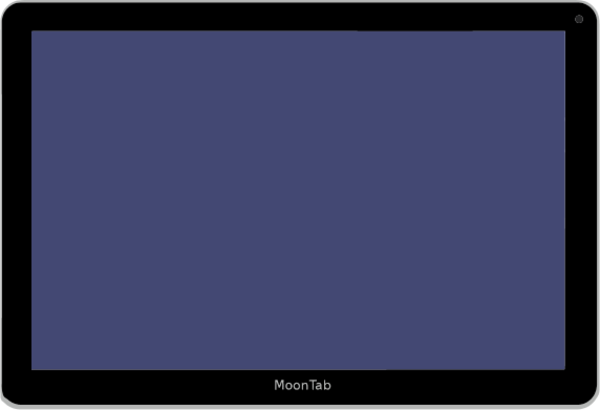}}
node[left=-0.1 of us3,font=\small,color=byzantium] {} node[below= 0cm of us3,font=\footnotesize] {$S_{n_1}$} node[left =0 of us3,font=\small,color=byzantium] {$\w_{n_1}$};;
 
\path[every node/.style={},color=black,->]
   (s.north west)    	edge[bend right]  node[midway,above left=0 and -0.4 cm,sloped,font=\footnotesize] {$\mathbf{u}+\mathbf{x}$} (us1) 
   (s.west)		edge[bend right]  node[midway,below=0,sloped,font=\footnotesize] {$\mathbf{u}+\mathbf{x}$}  (us2)
   (s.south west)		edge[bend left]  node[midway,below=0,sloped,font=\footnotesize] {$\mathbf{u}+\mathbf{x}$}   (us3)
			
   (us1)		edge[dashed]  node[midway,above =0, sloped,font=\footnotesize]  {$S_{1}(\mathbf{u}+\mathbf{x})$}  	(s.north west) 
   (us2)		edge[dashed]  node[midway,below =0, sloped,font=\footnotesize]  {$S_{2}(\mathbf{u}+\mathbf{x})$}  	(s)
   (us3)		edge[dashed]  node[midway,below =0, sloped,font=\footnotesize]  {$S_{n_1}(\mathbf{u}+\mathbf{x})$}  	(s.south west)
   ;
   
\node[above=0 of us1,color=byzantium] {Group $1$};

\end{tikzpicture}}
\captionsetup{subtype,width=\textwidth}
\caption{$\m$ sends $\mathbf{x}+\mathbf{u}$ and  $\mathbf{u}$  to the workers of group 1 and 2, respectively. $\w_i$ computes $S_i\mathbf{x}$ and sends the result to $\m$.}
\label{fig:apdxfig2}
\end{minipage}
\caption{Secure distributed matrix multiplication with $n$ workers, where $\m$ needs to hide both $A$ and $\mathbf{x}$. $\m$ divides the workers intro two disjoint groups of cardinality $n_1$ and $n_2$, respectively, such that $n_1+n_2=n$. Now $\m$ deals with the groups as two separate $(n_1,k_1,z_1)$ and $(n_2,k_2,z_2)$ systems, where $z_1<k_1<n_1$ and $z_2<k_2<n_2$. To hide $\mathbf{x}$, $\m$ generates a random vector $\mathbf{u}$ and sends $\mathbf{x}+\mathbf{u}$ to group 1 and $\mathbf{u}$ to group 2. Hence, $\m$ decodes $A\mathbf{x}$ after decoding $A\mathbf{u}$ and $A(\mathbf{x}+\mathbf{u})$.}\label{fig:apdx1}
\end{figure}
{Throughout the paper we assumed privacy over one iteration, i.e., the Master needs to hide only $A$. In the following we describe how our scheme can be generalized to achieve privacy over the whole algorithm, i.e., the Master needs to hide $A$ and the attribute vectors $\mathbf{x}^1,\mathbf{x}^2,\dots$. Since the algorithm is iterative, we focus on one iteration and the remaining follows similarly. Let $A$ be an $m\times \ell$ matrix and $\mathbf{x}$ be an $\ell \times 1$ vector that the Master $\m$ wishes to distributively multiply. Let $n$ be the number of workers $\w_i$, $i=1,\dots, n,$ that volunteer to help $\m$. The idea is to divide the workers into two disjoint groups and ask each of them to securely multiply $A$ by a vector that is statistically independent of $\mathbf{x}$. Then, the Master decodes $A\mathbf{x}$ from the results of both multiplications, as described next. $\m$ divides the workers into two groups of cardinality $n_1$ and $n_2$ such that $n_1+n_2=n$. Afterwards, $\m$ chooses $z_1<k_1<n_1$ and $z_2<k_2<n_2$ and encodes $A$ into $n$ shares using an $(n_1,k_1,z_1)$ and an $(n_2,k_2,z_2)$ Staircase codes (or classical secret sharing codes). Thereafter, $\m$ distributes the shares to the workers such that the groups form two disjoint $(n_1,k_1,z_1)$ and $(n_2,k_2,z_2)$ systems. To hide $\mathbf{x}$, $\m$ generates a random vector $\mathbf{u}$ of same size as $\mathbf{x}$ and sends $\mathbf{x}+\mathbf{u}$ to the first group and $\mathbf{u}$ to the second group. According to our scheme, $\m$ decodes $A(\mathbf{x}+\mathbf{u})$ and $A\mathbf{u}$ after receiving enough responses from the workers of each group. Hence, $\m$ can decode $A\mathbf{x}$. Note that no information about $\mathbf{x}$ is revealed because it is one-time padded by $\mathbf{u}$. We illustrate the idea in Figure~\ref{fig:apdx1}.}

\bibliographystyle{ieeetr}
\bibliography{Rawad}

\end{document}